\documentclass[sigconf]{acmart}
\pdfoutput=1
\AtBeginDocument{%
  \providecommand\BibTeX{{%
    \normalfont B\kern-0.5em{\scshape i\kern-0.25em b}\kern-0.8em\TeX}}}

\setcopyright{acmcopyright}
\copyrightyear{2020}
\acmYear{2020}
\setcopyright{acmcopyright}\acmConference[MM '20]{Proceedings of the 28th ACM International Conference on Multimedia}{October 12--16, 2020}{Seattle, WA, USA}
\acmBooktitle{Proceedings of the 28th ACM International Conference on Multimedia (MM '20), October 12--16, 2020, Seattle, WA, USA}
\acmPrice{15.00}
\acmDOI{10.1145/3394171.3413976}
\acmISBN{978-1-4503-7988-5/20/10}

\usepackage{graphicx}
\usepackage{amsmath}
\usepackage{amssymb}
\usepackage{algorithm}
\usepackage{algorithmic}
\usepackage{multirow}
\usepackage{setspace}
\usepackage{diagbox}
\usepackage{fancyhdr}

\begin{document}
\fancyhead{}
\title{Adv-watermark: A Novel Watermark Perturbation for Adversarial Examples}

\author{Xiaojun Jia}
\email{jiaxiaojun@iie.ac.cn}
\affiliation{%
  \institution{Institute of Information Engineering, Chinese Academy of Sciences}
  \state{Beijing}
  \country{China}
}

\author{Xingxing Wei}
\authornote{Corresponding Author}
\email{xxwei@buaa.edu.cn}
\affiliation{%
 \institution{Beijing Key Laboratory of Digital Media, School of Computer Science and Engineering, Beihang University}
  \state{Beijing}
  \country{China}
 }

\author{Xiaochun Cao}
\authornotemark[1]
\email{caoxiaochun@iie.ac.cn}
\affiliation{%
  \institution{Institute of Information Engineering, Chinese Academy of Sciences}
  \state{Beijing}
  \country{China}
}

\author{Xiaoguang Han}
\email{hanxiaoguang@cuhk.edu.cn}
\affiliation{%
  \institution{Shenzhen Research Institute of Big Data, the Chinese University of Hong Kong (Shenzhen)}
  \city{Shenzhen}
  \state{Guangdong}
  \country{China}
}

\renewcommand{\shortauthors}{Jia, et al.}

\begin{abstract}
Recent research has demonstrated that adding some imperceptible perturbations to original images can fool deep learning models. However, the current adversarial perturbations are usually shown in the form of noises, and thus have no practical meaning. Image watermark
is a technique widely used for copyright protection. We can regard image watermark as a kind of meaningful noises and adding it to the original image will not affect people's understanding of the image content, and will not arouse people's suspicion. Therefore, it will be interesting to generate adversarial examples using watermarks. In this paper, we propose a novel watermark perturbation for adversarial examples (Adv-watermark) which combines image watermarking
techniques and adversarial example algorithms. Adding
a meaningful watermark to the clean images can attack the DNN models. Specifically, we
propose a novel optimization algorithm, which is called Basin Hopping Evolution (BHE), to generate adversarial watermarks in the black-box attack mode. Thanks to the BHE, Adv-watermark only requires a few queries from
the threat models to finish the attacks. A series of experiments conducted on
ImageNet and CASIA-WebFace datasets show that the proposed method can efficiently generate adversarial examples, and outperforms the
state-of-the-art attack methods. Moreover, Adv-watermark
is more robust against image transformation defense methods.

\end{abstract}




\maketitle
\begin{figure}[tt]
\begin{center}
   \includegraphics[width=1\linewidth]{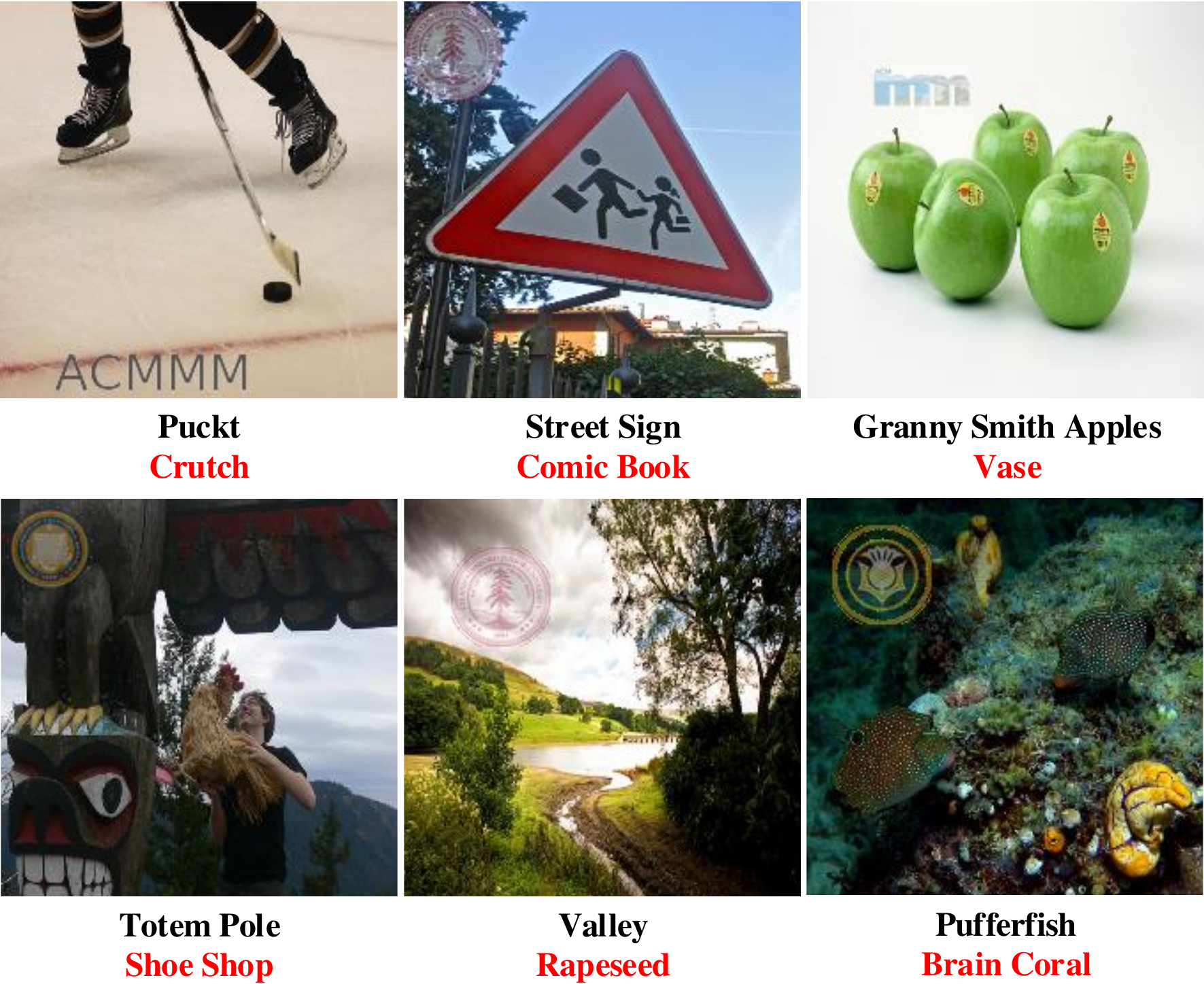}
\end{center}
   \caption{Adversarial examples with watermark perturbations.
   The original class labels are in black text and the adversarial class labels  are in red text.
   }
\label{fig:cvpr1}
\vspace{-.3cm}
\end{figure}

\section{Introduction}
Recent literature has found that Deep Neural Networks (DNNs) are vulnerable to the
adversarial examples which are generated by adding some imperceptible noises to the  clean images
\cite{goodfellow2014generative}. Generally speaking, attack methods can be divided into two categories:
white-box attack methods and black-box attack methods.
The white-box attack \cite{goodfellow2014explaining, kurakin2016adversarial, moosavi2016deepfool, carlini2017towards}
denotes that the attacker has complete access to the target model such
as model parameters, model structure, etc. And the black-box attack \cite{su2019one, schott2018adversarially, engstrom2019exploring,wiel2017decisionbased, liang2020efficient} denotes that
the attacker can only access the output of the target model. The above methods achieve attacks by generating imperceptible
perturbations. They use $L_{0}, L_{2}, L_{\infty}$ to bound the noises.
Recently, more and more researchers pay attention to generating
realistic adversarial examples without the $L_{p}$ norm limitation \cite{qiu2019semanticadv, lagae2010survey, engstrom2019exploring}.

\par Watermarking  methods \cite{hartung1999multimedia} play an important role in protecting intellectual property rights.
It embeds some specific information of the copyright holder (such as university logos, ownership descriptions, etc)
into the multimedia data according to the requirements of users.
In \cite{mintzer1997effective}, Mintzer et al. describe the characteristics of visible watermarks.
The visible watermark should be visible but does not significantly obscure the details of the host image.
\par In this paper, we propose a novel adversarial attack which generates
adversarial examples using watermarks. We find that although watermarks
do not affect people's understanding of the image content, and adding specific watermarks
to the clean images can fool the DNN models. The specific watermarks
refer to the specific position and transparency of them.
We mainly consider using visible watermarks to generate adversarial examples.
In detail, we use alpha blending \cite{shen1998dct} to achieve watermark embedding.
The host image and the watermark are multiplied by a scaling factor.
The scaling factor is manipulated in the $\alpha$ channel of the image,
which decides the image's transparency.
\par As for a certain watermark, the DNN models can be successfully attacked only
by adding the watermark with the specific transparency to a specific position of the
host image.
Considering this,
we propose a novel attack method to generate watermark adversarial perturbations. Specifically, we propose a Basin Hopping Evolution (BHE) algorithm to find the appropriate transparency of the watermark image and
the appropriate position within the host image to embed watermark. BHE is proposed based on the Hopping Evolution (BH) \cite{wales1997global}, where we find it usually falls into a local optimum and
fails in attacking DNN models. In contrast, BHE has multiple
initial starting points and crossover operation to
keep the diversity of solutions. In this way,
BHE makes it easier to find a global optimal solution and thus
achieves a higher attack success rate than BH.
The proposed method achieves attacks by using a little information (predicted probability of the classification model).
It does not need the inner information of DNNs such as
 network structures and weights.
Therefore, it belongs to the black-box attack.

\par
\begin{figure}[tt]
\begin{center}
   \includegraphics[width=0.95\linewidth]{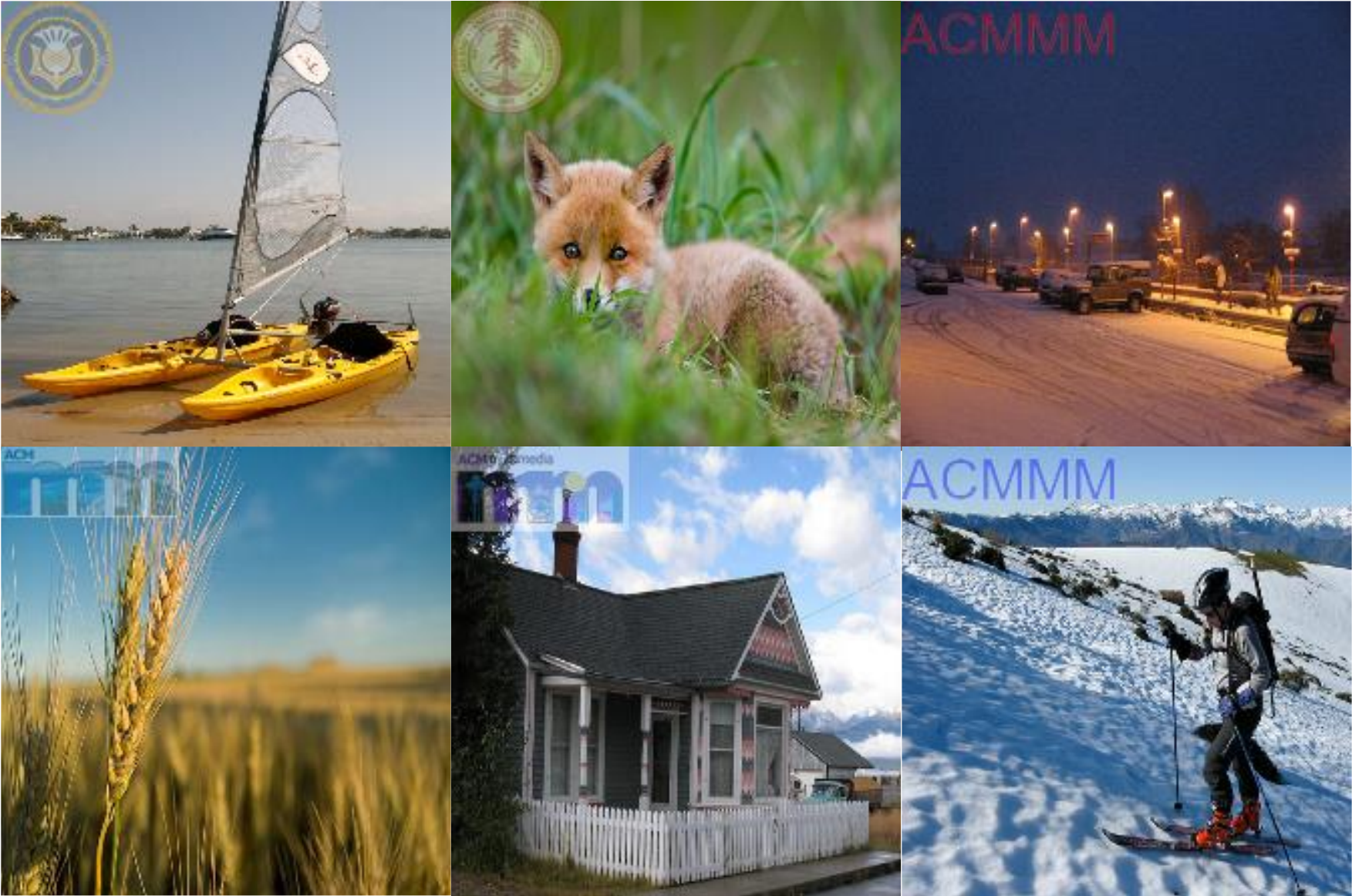}
\end{center}
   \caption{In this paper, we explore two kinds of media as the watermark: logos and texts.
   These six host images are randomly selected from ImageNet.
   }
\label{fig:cvpr2}
\vspace{-.5cm}
\end{figure}

 Besides the ability to perform adversarial attacks, Adv-watermark also inherits the function of the visible watermark. That's to say, Adv-watermark can also protect the copyright of the image because it carries the owner's description. Therefore, Adv-watermark can accomplish two functions at the same time. This is a major advantage compared with the previous research. Specifically, peoples tend to share their images on social media to record their lives. They usually add a visible watermark to protect their copyright. But their images can also be identified and embezzled by malicious software. Adv-watermark can be used to avoid this situation. It not only protects the copyright of the image but also performs adversarial attacks to avoid being embezzled by malicious software. In this paper, we explore two kinds of media as the watermarks: logos and texts. Figure \ref{fig:cvpr2} lists the used watermarks, and some generated Adv-watermark examples are shown in Figure \ref{fig:cvpr1}.

\par In summary, this paper has the following contributions:

\par 1) We propose the Adv-watermark, a novel watermark perturbation for adversarial examples,
which combines image watermarking techniques and adversarial example algorithms.
Compared with the previous works, the proposed adversarial example is more
realistic and effective. 

\par 2) We propose a novel optimization algorithm, which
is called Basin Hopping Evolution (BHE), to generate
adversarial examples efficiently. The proposed method
adopts a population-based global search strategy to generate
adversarial examples, and can achieve
high performance in attacking DNN models.

\par 3) Compared with the previous black-box attack methods, the proposed method can achieve a
higher attack success rate. Moreover, the state-of-the-art image transformation defense methods can not defend the proposed attack method. The code is released at https://github.com/jiaxiaojunQAQ/Adv-watermark.git.
\par The remainder of this paper is organized as follows. Section 2 briefly reviews the related work.
Section 3 introduces the details of the proposed Adv-watermark.
Section 4 shows a series of experimental results and analysis. Finally, Section 5 gives the
conclusion.

\section{Related work}
In this section, we investigate the attack methods and the visible watermarking methods.

\subsection{Attack methods}
In \cite{goodfellow2014explaining}, Goodfellow et al. devise an effective
method to calculate the adversarial examples, and the adversarial perturbation is generated according
to the direction of the gradient change of the DNNs.
This method is also called FGSM. Iterative FGSM (I-FGSM) \cite{kurakin2016adversarial} is an improved
version of FGSM. I-FGSM constructs an adversarial example
by multi-step and smaller movements, which greatly improves the success rate of the attack.
The most common adversarial attack methods are under the $L_{\infty}$ and $L_{2}$ distance metric.
But in \cite{papernot2016limitations}, Papernot et al. propose to build adversarial saliency maps
to generate adversarial examples under $L_{0}$ norm.
Moosavi-Dezfooli et al. propose a simple and accurate method (Deepfool) \cite{moosavi2016deepfool} to
efficiently generate the adversarial examples. Moreover, they further propose the universal perturbation
based on Deepfool in \cite{moosavi2017universal}. And in \cite{carlini2017towards}, Carlini and Wagner
propose three attack methods to attack defensive distillation Networks \cite{papernot2016distillation}.
In \cite{su2019one}, Su at el. propose to generate one-pixel adversarial perturbations based on differential evolution (DE).

\subsection{Visible watermarking methods}
In \cite{kankanhalli1999adaptive}, Kankanhalli et al.
propose a visible watermarking technique that can find
the strength of the watermark image and the location of the
host image.
In \cite{shen1998dct}, Shen et al. propose to use the alpha blending technique to generate the visible watermark.
A removable visible watermark is proposed in \cite{hu2005algorithm}. They design a vision watermarking
algorithm suitable for the different requirements of the
applications. In \cite{liu2010generic}, Liu et at. propose
a new approach to generate a generic lossless visible watermark. The proposed method makes use of deterministic one-to-one mappings of image pixel values to achieve
generating the visible watermark. In \cite{huang2006contrast}, Huang et al. design a visible
watermarking algorithm for digital right management.
A contrast-sensitive function and block classification
are used to achieve a better visual effect in the discrete wavelet transform domain.

\section{Methodology}

In this section, we introduce the proposed method from three
aspects: visible watermarking, problem formulation and
problem solving.

\begin{figure}[tt]
\begin{center}
   \includegraphics[width=0.95\linewidth]{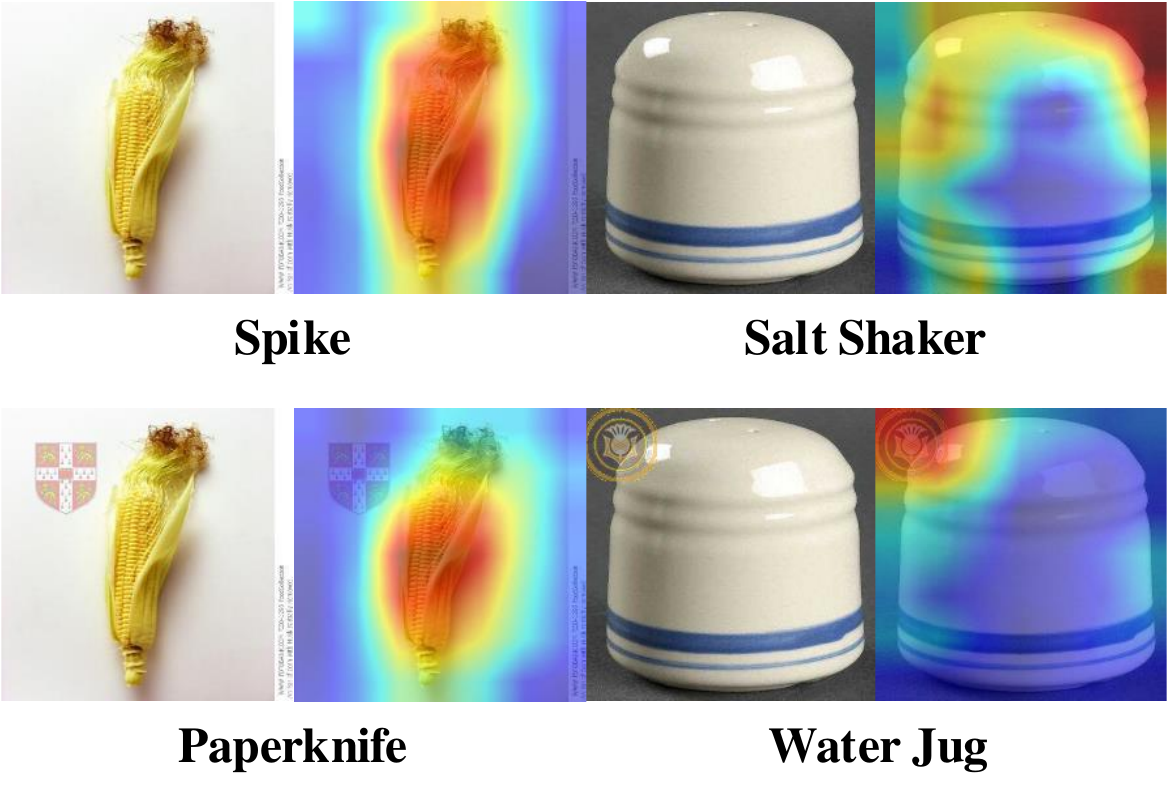}
\end{center}
   \caption{ The top row is the original images (they are correctly classified by Resnet101) and
   their corresponding heat-maps (generated by Grad-CAM algorithm).The bottom row is
   the adversarial images with the visible watermark and their corresponding heat-maps.
   The image classification labels are in black color.}
\label{fig:cvpr4}
\vspace{-.3cm}
\end{figure}

\subsection{Visible Watermarking}
We use alpha blending in \cite{shen1998dct} to generate a visible watermark.
Alpha channel($\alpha$ channel) refers to the transparency of a foreground
region w.r.t. the background image.
In this paper, we use $\alpha$ to represent the value of
the alpha channel, $H$ to represent the host image whose size is $N \times M$, $W$ to
represent the watermark image whose size is $n \times m$ and $G$ to represent the
generated image with a watermark whose size is $N \times M$. When $i\in (p,p+n), j\in (q,q+m)$,
the generation for $G$ is formulated as:
\begin{equation}
v(G)_{i,j} =(v(W)_{i-p,j-q}*\alpha +v(H)_{i,j}*(255-\alpha))/255
\end{equation}
 when $i\not\in (p,p+n), j\not\in (q,q+m)$,
 $G$ is formulated as:
\begin{equation}
v(G)_{i,j} =v(H)_{i,j},
\end{equation}
where $v(x)$ denotes the image $x$, the subscript
$i,j$ of $v(x)$ represent the pixel position, and
$ p,q $ represent the position where the watermark image
is embedded.
As for the image watermark, we use UC Berkeley, CMU, MIT, Cambridge
and Stanford University logo watermarks. Simultaneously,
we also use the official ACMMM logo from 2016 to 2020.
As for text watermark, we use red, green, blue, black
and gray fonts to generate adversarial examples.
We also synthesize watermark images in different sizes to explore scale-ware effects.
It is formulated as:
\begin{equation}
\begin{split}
& \eta =\min ((W_{h} \ast sl)/W_{w}, (H_{h} \ast sl) / H_{w}), \\
& W_{sw}=W_{w}\ast \eta, H_{sw}=H_{w} \ast \eta,
\end{split}
\end{equation}
where $W_{h}$ and $H_{h}$ represent the width and height of the host image.
 $W_{w}$ and $H_{w}$ represent the width and height of the watermark image.
$sl$ is the scaling factor. And $W_{sw}$ and $H_{sw}$ represent the width
and height of the scaled watermark image. Note that in this paper,
we focus on the position and transparency of the watermark, not the rotation, etc.

\subsection{Problem Formulation}
We disguise adversarial noise as a visible watermark to achieve stealthiness.
And the generation of adversarial examples is only related to
the position and transparency of the watermark.
Generating adversarial
watermark images can be formalized as an optimization problem with constraints. The host
image is assumed as $H$, the well-trained classification model is assumed as $f$ and the
correct classification class of $H$ is $t$. $f_{t}(H)$ is the probability of $H$
belonging to the class $t$. Simultaneously, let $W$ be the watermark image and $g(H,W,p,q, \alpha)$
be the visible watermark algorithm. It embeds the watermark image $W$ in the position $(p,q)$ of the
host image $H$. The $p$, $q$ and $\alpha$ are dependent on $W$, $H$, $f$. And the limitation of maximum transparency of the watermark is $L$. In the case of untargeted attacks, the goal of
generation of adversarial examples can be transformed into finding the optimized solution $e(p, q, \alpha)^{\ast}$.
It is formulated as:
\begin{equation}
\begin{array}{cl}{\underset{e(p, q, \alpha)^{*}}{\operatorname{minimize}}} & {f_{t}(g(H,W,p,q,\alpha))} \\
{\text { subject to }} & {\mathbf{\alpha} \leq L}\end{array}
\end{equation}
\par This problem involves two values: 1) the position $(p,q)$ of the watermark in the host image and
2) the transparency $\alpha$ of the watermark.
Embedding the adversarial watermark which can be regarded as a practical perturbation
into the host image modifies the local information of the host image.
In this way, the adversarial watermark perturbation permits
a clean image to be an adversarial example.
Without affecting the visual effect of the image, the adversarial
watermark disturbs the important local regions which
determine the image classification to attack the
well-trained classification model.
This is illustrated in Figure \ref{fig:cvpr4}.
From the heat-maps which are generated by Gradient-weighted Class Activation Mapping (Grad-CAM) \cite{Selvaraju_2017_ICCV},
it is clear why the Resnet101 predicts the input images
as the corresponding correct classes. And embedding the adversarial watermark into the image can
modify the distribution of the maximum points on the generated heat-map.

\subsection{Problem Solving}

\begin{figure}[tt]
\begin{center}
   \includegraphics[width=1\linewidth]{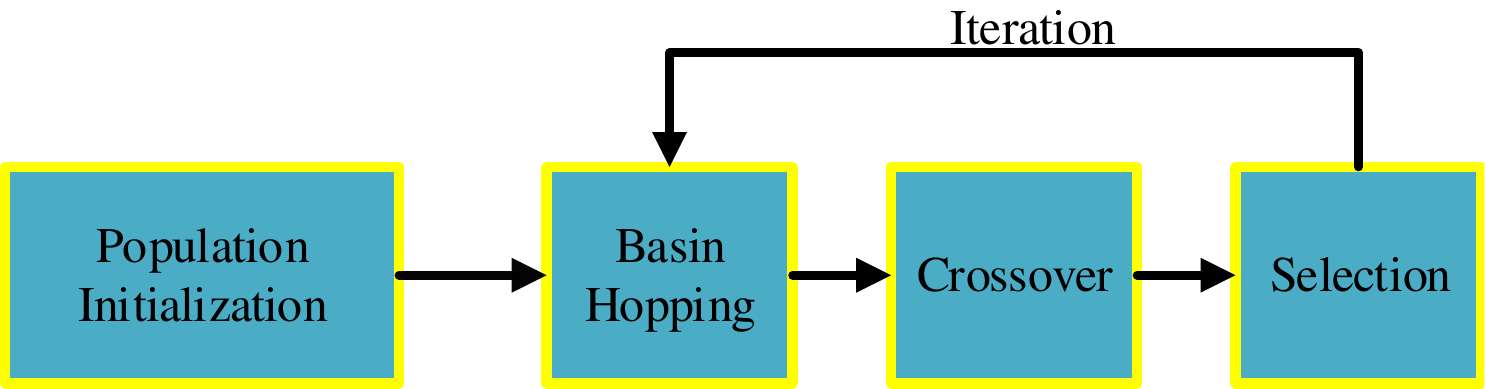}
\end{center}
\caption{ The process of Basin Hopping Evolution.}
\label{fig:BHE}
\end{figure}

We propose a novel optimization algorithm, which is called Basin
Hopping Evolution(BHE).
The proposed method is a heuristic random search algorithm based on Basin Hopping, which can be used for finding the global minimum of a multivariate function. As shown in Figure \ref{fig:BHE}, BHE includes Basin Hopping, crossover
and selection operations. During each iteration, the current solutions (parents)
use BH to produce a set of better solutions and conduct crossover operation to generate a new set of candidate solutions (children). And then in selection operation, compared with the corresponding parents to conduct, if the children are
more suitable for the current population evolution (posses the smaller multivariate function value), they survive and are passed to the
next generation.

\subsubsection{Population Initialization}
BHE is an optimization algorithm based on group evolution.
We regard each solution as an individual of a population.
And the elements ($p, q$ and $\alpha$)  are considered
as its genes. Let $X_{i,g}$ denote the $i$-th individual in the
$g$-th generation population. And $X_{i,g,j} (j=0,1,2)$ denotes the $j$-th gene
of $X_{i,g}$. Therefore, we initialize a population as follows:
\begin{equation}
\begin{split}
X_{i,0,j}=X_{\min,j}+rand(0,1)\cdot(X_{\max,j}-X_{\min, j}), j=0,1,2
\end{split}
\end{equation}
where  $X_{i,0,j}$ is the $j$-th gene of the $i$-th individual in the initial population , $X_{\min,j}$ is the minimum of the $j$-th gene and $X_{\max,j}$ is the maximum of the $j$-th gene.
\begin{figure}[tt]
\begin{center}
   \includegraphics[width=1\linewidth]{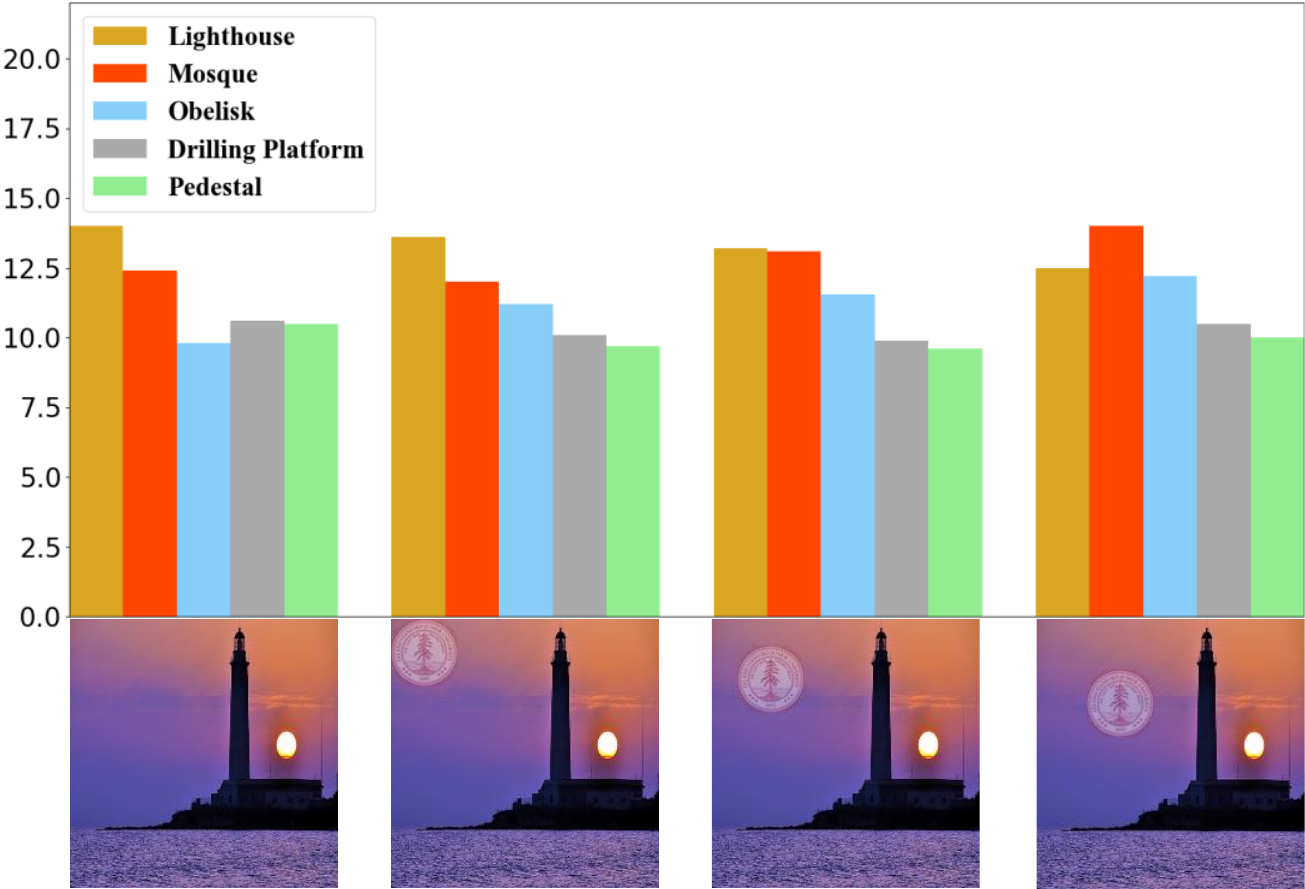}
\end{center}
\caption{ The histogram of the top 5 category  confidence of
   generated images in the generation process of the adversarial examples
   using BHE.}
\label{fig:cvpr5}
\vspace{-.3cm}
\end{figure}
\subsubsection{Basin Hopping}
Basin Hopping (BH) is a stochastic optimization algorithm.
During each iteration, BH generates some new coordinates with random perturbations, next finds
the local minimization, and finally accepts or rejects
the new coordinates according to the minimized function
value. We use BH to evolve a better individual $V_{i,g}$ from $X_{i,g}$.
\par In detail, $f_{t}(g(H,W,p,q,\alpha))$ is assumed as $f_{t}(\cdot)$.
Starting with $X_{i,g}$, a local optimal solution $V_{i,g}$
of the function $f_{t}(\cdot)$ is found by using a minimization method $L(\cdot)$.
Next we start the global search iterations and use $\mu_{g}(X_{i,g})$ to represent the
global neighborhood of $X_{i,g}$. It is formulated as:
\begin{equation}
\mu_{g}(X_{i,g})=[X_{i,g}, X_{i,g}+r*\vec{d}],
\end{equation}
where $\vec{d}$ is an n-dimensional Gaussian$(0, 1)$ variable and
$r$ is a fixed step size.
A new starting point is selected from the global neighborhood of
$V_{i,g}$. It is stored as $V_{i,g}$.It is formulated as:
\begin{equation}
V_{i,g}=G(\mu_{g}(V_{i,g}))
\end{equation}
And then starting with $V_{i,g}$,
a local search is performed and the result is stored as $S_{i,g}$.
Finally, we use a function $Accept(V_{i,g},S_{i,g})$ to choose $V_{i,g}$ or $S_{i,g}$. And it is formulated as:
\begin{equation}
Accept(V_{i,g},S_{i,g})=
\begin{cases}
1& f_{t}(S_{i,g}) \leq f_{t}(V_{i,g})\\
0& f_{t}(S_{i,g}) > f_{t}(V_{i,g})
\end{cases}
\end{equation}
The detail description is given in Algorithm \ref{alg:Framwork}.
To represent BH algorithm simplify, it can be formulated
as:
\begin{equation}
V_{i,g}=BH(X_{i,g}, I),
\end{equation}
where $X_{i,g}$ represents the $i$-th solution in the $g$-th generation population,
$V_{i,g}$ represents the corresponding better solution using BH, $BH(\cdot)$ represents
the BH algorithm and $I$ indicates the maximum number of Basin Hopping iterations which is
a super parameter which we use a large number of experiments to certify.
\begin{algorithm}[htb]
\caption{BH algorithm}
\label{alg:Framwork}
\begin{algorithmic}[1]
\REQUIRE
The watermark image $W$, the host image $H$, the well-trained classifier $f$
and $X_{i,g}$
\ENSURE  $V_{i,g}$

\STATE $V_{i,g}=L(f_{t}(\cdot), X_{i,g})$;
\REPEAT
\STATE $V_{i,g}=G(\mu_{g}(V_{i,g}))$;
\STATE $S_{i,g}=L(f_{t}(\cdot), V_{i,g})$;

\IF {$Accept(V_{i,g}, S_{i,g})$}
\STATE $V_{i,g}=S_{i,g}$;
\ENDIF

\UNTIL{global stopping rule is satisfied}

\RETURN $V_{i,g}$
\end{algorithmic}
\end{algorithm}
\subsubsection{Crossover}
As for the current solution (parents) $X_{i,g}$ and the corresponding BH
optimization solution $V_{i,g}$, we conduct crossover operation
to get a candidate solution (child) $U_{i,g}$. It is formulated as:
\begin{equation}
U_{i, g, j}=\left\{\begin{array}{l}V_{i, g, j}, r a n d(0,1) \leq \text {CR } \\
X_{i, g, j}, \text { others}\end{array}\right.
\end{equation}
where $U_{i, g, j}$ is the $j$-th gene of  $U_{i, g}$,
$V_{i, g, j}$ is the $j$-th gene of  $V_{i, g}$,
$X_{i, g, j}$ is the $j$-th gene of  $X_{i, g}$
and $\text {CR}$ is the crossover probability
which represents the degree of information exchange
in the population evolution. It is
a super parameter which we use a large number of experiments to certify.

\subsubsection{Selection}
We adopt a greedy selection strategy to select a better
solution as the next generation solution.
It is formulated as:
\begin{equation}
X_{i, g+1}=\left\{\begin{array}{l}\mathrm{U}_{i, g}, f_{t}\left(U_{i, g}\right) \leq f_{t}\left(X_{i, g}\right) \\ X_{i, g} \text { others }\end{array}\right.
\end{equation}
\par The detail description of BHE is given in Algorithm \ref{alg:Framwork1}. And the generation process
of the adversarial examples by using BHE is shown in Figure \ref{fig:cvpr5}.

\begin{algorithm}[htb]
\caption{BHE algorithm }
\label{alg:Framwork1}
\begin{algorithmic}[1]
\REQUIRE
Population: $M$; Dimension: 3; Generation: $N$; Iteration: $I$;

\ENSURE  The best solution -$\bigtriangleup$
\STATE $g \leftarrow 0$;
\FOR{$i=1$ to $M $}
\FOR{$j=1$ to $3 $}
\STATE  $X_{i,0,j}=X_{\min, j}+rand(0,1)\cdot(X_{\max, j}-X_{\min, j})$
\ENDFOR
\ENDFOR
\WHILE{$f_{t}(\bigtriangleup)\geq\varepsilon$ and $g\leq N$}
\FOR{$i=1$ to $M $}
\STATE $\blacktriangleright$ Basin Hopping
\STATE  $V_{i,g}=BH(X_{i,g}, I)$
\STATE $\blacktriangleright$ Crossover
\FOR{$j=1$ to $3 $}
\STATE  $U_{i,g,j}=Crossover(V_{i,g,j}, X_{i,g,j})$
\ENDFOR
\STATE $\blacktriangleright$ Selection
\IF{$f_{t}(U_{i,g}) \leq f_{t}(X_{i,g})$}
\STATE $X_{i,g}=U_{i,g}$
\IF{$f_{t}(X_{i,g}) \leq f_{t}(\bigtriangleup)$}
\STATE $\bigtriangleup=X_{i,g}$
\ENDIF
\ELSE
\STATE $X_{i,g}=X_{i,g}$
\ENDIF
\ENDFOR
\STATE $g \leftarrow g+1$
\ENDWHILE

\end{algorithmic}
\end{algorithm}

\begin{table}[tt]
\begin{center}
\caption{ Selection of hyper-parameters in BHE }
\label{table:cvprtable00}
\setlength{\tabcolsep}{1mm}{
\begin{tabular}{|c|c|c|c|c|c|c|c|}
\hline
        & $ I=2$     & $ I=3$     & $ I=4 $    & $ I=5$     & $ I=6 $    & Average \\ \hline
CR=0.5  & 59.3\% & 58.1\% & 58.1\% & \textbf{60.0\%} & 58.1\% & 58.4\%  \\ \hline
CR=0.6  & 56.2\% & 57.5\% & 57.5\% & 59.3\% & 58.7\% & 57.6\%  \\ \hline
CR=0.7  & 58.1\% & 58.7\% & 59.3\% & 59.3\% & \textbf{60.0\%} & 58.7\%  \\ \hline
CR=0.8  & 58.7\% & 56.8\% & \textbf{60.0\%} & \textbf{60.0\%} & \textbf{60.0\%} & 58.7\%  \\ \hline
CR=0.9  & 58.1\% & \textbf{60.0\%} & 59.7\% & 59.3\% & 59.3\% & 58.9\%  \\ \hline
CR=1.0  & 59.3\% & 59.3\% & \textbf{60.0\%} & \textbf{60.0\%} & \textbf{60.0\%} & 59.3\%  \\ \hline
Average & 58.3\% & 58.4\% & 58.9\% & 59.6\% & 59.3\% & 58.6\%  \\ \hline
\end{tabular}
}
\end{center}
\end{table}
\vspace{-.4cm}

\begin{table*}[tt]
\caption{The attack success rates of individual logo or text watermark. }
\label{table:cvprtable01}
\vspace{-.4cm}
\begin{center}

\begin{tabular}{|c|c|l|c|l|c|l|c|l|c|l|c|l|}
\hline
\multicolumn{13}{|c|}{ACMMM logo watermarks}                                                                                                                                                                           \\ \hline
          & \multicolumn{2}{c|}{Alexnet}   & \multicolumn{2}{c|}{VGG19}     & \multicolumn{2}{c|}{SqueezeNet1\_0} & \multicolumn{2}{c|}{Resnet101} & \multicolumn{2}{c|}{InceptionV3} & \multicolumn{2}{c|}{Average}   \\ \hline
scale=2/3 & \multicolumn{2}{c|}{88\%/\textbf{92\%}} & \multicolumn{2}{c|}{77\%/\textbf{83\%}} & \multicolumn{2}{c|}{85\%/\textbf{88\%}}      & \multicolumn{2}{c|}{78\%/\textbf{83\%}} & \multicolumn{2}{c|}{77\%/\textbf{79\%}}   & \multicolumn{2}{c|}{81\%/\textbf{85\%}} \\ \hline
scale=1/2 & \multicolumn{2}{c|}{80\%/\textbf{88\%}} & \multicolumn{2}{c|}{69\%/\textbf{80\%}} & \multicolumn{2}{c|}{76\%/\textbf{82\%}}      & \multicolumn{2}{c|}{70\%/\textbf{78\%}} & \multicolumn{2}{c|}{65\%/\textbf{74\%}}   & \multicolumn{2}{c|}{72\%/\textbf{80\%}} \\ \hline
scale=1/3 & \multicolumn{2}{c|}{68\%/\textbf{76\%}} & \multicolumn{2}{c|}{54\%/\textbf{68\%}} & \multicolumn{2}{c|}{56\%/\textbf{69\%}}      & \multicolumn{2}{c|}{56\%/\textbf{66\%}} & \multicolumn{2}{c|}{51\%/\textbf{61\%}}   & \multicolumn{2}{c|}{57\%/\textbf{68\%}} \\ \hline
scale=1/4 & \multicolumn{2}{c|}{58\%/\textbf{69\%}} & \multicolumn{2}{c|}{43\%/\textbf{59\%}} & \multicolumn{2}{c|}{46\%/\textbf{62\%}}      & \multicolumn{2}{c|}{47\%/\textbf{58\%}} & \multicolumn{2}{c|}{41\%/\textbf{52\%}}   & \multicolumn{2}{c|}{47\%/\textbf{60\%}} \\ \hline
Average   & \multicolumn{2}{c|}{74\%/\textbf{81\%}} & \multicolumn{2}{c|}{61\%/\textbf{72\%}} & \multicolumn{2}{c|}{66\%/\textbf{75\%}}      & \multicolumn{2}{c|}{63\%/\textbf{71\%}} & \multicolumn{2}{c|}{59\%/\textbf{62\%}}   & \multicolumn{2}{c|}{65\%/\textbf{73\%}} \\ \hline
\multicolumn{13}{|c|}{University logo watermarks}                                                                                                                                                                      \\ \hline
          & \multicolumn{2}{c|}{Alexnet}   & \multicolumn{2}{c|}{VGG19}     & \multicolumn{2}{c|}{SqueezeNet1\_0} & \multicolumn{2}{c|}{Resnet101} & \multicolumn{2}{c|}{InceptionV3} & \multicolumn{2}{c|}{Average}   \\ \hline
scale=2/3 & \multicolumn{2}{c|}{96\%/\textbf{98\%}} & \multicolumn{2}{c|}{96\%/\textbf{96\%}} & \multicolumn{2}{c|}{95\%/\textbf{97\%}}      & \multicolumn{2}{c|}{96\%/\textbf{97\%}} & \multicolumn{2}{c|}{96\%/\textbf{98\%}}   & \multicolumn{2}{c|}{96\%/\textbf{97\%}} \\ \hline
scale=1/2 & \multicolumn{2}{c|}{90\%/\textbf{95\%}} & \multicolumn{2}{c|}{88\%/\textbf{90\%}} & \multicolumn{2}{c|}{88\%/\textbf{91\%}}      & \multicolumn{2}{c|}{88\%/\textbf{90\%}} & \multicolumn{2}{c|}{87\%/\textbf{91\%}}   & \multicolumn{2}{c|}{89\%/\textbf{92\%}} \\ \hline
scale=1/3 & \multicolumn{2}{c|}{78\%/\textbf{88\%}} & \multicolumn{2}{c|}{74\%/\textbf{76\%}} & \multicolumn{2}{c|}{73\%/\textbf{79\%}}      & \multicolumn{2}{c|}{72\%/\textbf{76\%}} & \multicolumn{2}{c|}{68\%/\textbf{77\%}}   & \multicolumn{2}{c|}{73\%/\textbf{79\%}} \\ \hline
scale=1/4 & \multicolumn{2}{c|}{66\%/\textbf{78\%}} & \multicolumn{2}{c|}{62\%/\textbf{66\%}} & \multicolumn{2}{c|}{61\%/\textbf{71\%}}      & \multicolumn{2}{c|}{60\%/\textbf{66\%}} & \multicolumn{2}{c|}{54\%/\textbf{63\%}}   & \multicolumn{2}{c|}{61\%/\textbf{69\%}} \\ \hline
Average   & \multicolumn{2}{c|}{83\%/\textbf{90\%}} & \multicolumn{2}{c|}{80\%/\textbf{82\%}} & \multicolumn{2}{c|}{80\%/\textbf{84\%}}      & \multicolumn{2}{c|}{79\%/\textbf{82\%}} & \multicolumn{2}{c|}{76\%/\textbf{82\%}}  & \multicolumn{2}{c|}{80\%/\textbf{84\%}} \\ \hline
\multicolumn{13}{|c|}{Text watermarks}                                                                                                                                                                                 \\ \hline
          & \multicolumn{2}{c|}{Alexnet}   & \multicolumn{2}{c|}{VGG19}     & \multicolumn{2}{c|}{SqueezeNet1\_0} & \multicolumn{2}{c|}{Resnet101} & \multicolumn{2}{c|}{InceptionV3} & \multicolumn{2}{c|}{Average}   \\ \hline
font size=40   & \multicolumn{2}{c|}{89\%/\textbf{91\%}} & \multicolumn{2}{c|}{\textbf{82\%}/81\%} & \multicolumn{2}{c|}{84\%/\textbf{85\%}}      & \multicolumn{2}{c|}{74\%/\textbf{76\%}} & \multicolumn{2}{c|}{68\%/\textbf{73\%}}   & \multicolumn{2}{c|}{79\%/\textbf{81\%}} \\ \hline
font size=36   & \multicolumn{2}{c|}{85\%/\textbf{89\%}} & \multicolumn{2}{c|}{\textbf{79\%}/78\%} & \multicolumn{2}{c|}{80\%/\textbf{83\%}}      & \multicolumn{2}{c|}{69\%/\textbf{73\%}} & \multicolumn{2}{c|}{63\%/\textbf{69\%}}   & \multicolumn{2}{c|}{75\%/\textbf{78\%}} \\ \hline
font size=32   & \multicolumn{2}{c|}{82\%/\textbf{85\%}} & \multicolumn{2}{c|}{75\%/\textbf{76\%}} & \multicolumn{2}{c|}{76\%/\textbf{80\%}}      & \multicolumn{2}{c|}{65\%/\textbf{69\%}} & \multicolumn{2}{c|}{58\%/\textbf{65\%}}   & \multicolumn{2}{c|}{71\%/\textbf{75\%}} \\ \hline
font size=28   & \multicolumn{2}{c|}{75\%/\textbf{80\%}} & \multicolumn{2}{c|}{70\%/\textbf{71\%}} & \multicolumn{2}{c|}{71\%/\textbf{75\%}}      & \multicolumn{2}{c|}{59\%/\textbf{66\%}} & \multicolumn{2}{c|}{53\%/\textbf{60\%}}   & \multicolumn{2}{c|}{66\%/\textbf{70\%}} \\ \hline
Average   & \multicolumn{2}{c|}{83\%/\textbf{86\%}} & \multicolumn{2}{c|}{76\%/\textbf{76\%}} & \multicolumn{2}{c|}{78\%/\textbf{81\%}}      & \multicolumn{2}{c|}{67\%/\textbf{71\%}} & \multicolumn{2}{c|}{61\%/\textbf{66\%}}   & \multicolumn{2}{c|}{73\%/\textbf{76\%}} \\ \hline
\end{tabular}
\end{center}
\end{table*}

\begin{table*}[tt]
\caption{The attack success rates with limit of embedded watermark position }
\label{table:cvprtable02}
\vspace{-.4cm}
\begin{center}

\setlength{\tabcolsep}{6mm}{
\begin{tabular}{|c|c|c|c|c|c|}
\hline
 & scale=1/4                   & scale=1/5                   & scale=1/6                   & scale=1/7                   & scale=1/8                   \\ \hline
MIT logo       & {\color[HTML]{000000} 62\%} & {\color[HTML]{000000} 58\%} & {\color[HTML]{000000} 56\%} & {\color[HTML]{000000} 55\%} & {\color[HTML]{000000} 54\%} \\ \hline
ACMMM2020      & {\color[HTML]{000000} 63\%} & {\color[HTML]{000000} 59\%} & {\color[HTML]{000000} 58\%} & {\color[HTML]{000000} 57\%} & {\color[HTML]{000000} 53\%} \\ \hline
 & font size=22                & font size=21                & font size=20                & font size=19                & font size=18                \\ \hline
Red text       & {\color[HTML]{000000} 61\%} & {\color[HTML]{000000} 57\%} & {\color[HTML]{000000} 55\%} & {\color[HTML]{000000} 53\%}     & {\color[HTML]{000000} 50\%}     \\ \hline
\end{tabular}
}

\end{center}

\end{table*}

\section{Experimental results and analysis}

\subsection{Experiment Settings}
We conduct experiments based on ImageNet \cite{russakovsky2015imagenet}and CASIA-WebFace \cite{yi2014learning}.
In detail, we randomly select 1,000 images from them to conduct
the related experiments. We choose six classification models
with different structures as threat models: Alexnet \cite{krizhevsky2012imagenet}, VGG19
\cite{simonyan2014very}, SqueezeNet\cite{iandola2016squeezenet},
Resnet101 \cite{he2016deep}, InceptionV1 \cite{DBLP:conf/cvpr/SzegedyLJSRAEVR15} and
InceptionV3 \cite{szegedy2016rethinking}.
We also compare with other black-box attack methods to verify the
proposed method: spatial attack \cite{engstrom2019exploring},
boundary attack \cite{wiel2017decisionbased}, single-pixel attack \cite{su2019one} and
pointwise attack\cite{schott2018adversarially}. As for these attack methods,
we adopt their benchmark approaches and default parameters as recommended in
Foolbox \cite{rauber2017foolbox}.

\begin{table*}[tt]
\caption{ Comparison with other attack methods}
\label{table:cvprtable03}
\vspace{-.4cm}
\begin{center}
\setlength{\tabcolsep}{1mm}{
\begin{tabular}{|c|c|c|c|c|c|c|c|}
\hline
\diagbox{Network}{attacker}           & Spatial Attack & Boundary  Attack & Single-Pixel  & Pointwise Attack & SU logo                  & ACMMM2017           & Blue text                \\ \hline
Resnet101   & 52\%           & 37\%            & 5\%                 & 7\%              & 88\%  & 75\% & 73\% \\ \hline
InceptionV3 & 58\%           & 48\%            & 5\%                 & -                & 87\% & 72\% & 67\% \\ \hline
\end{tabular}
}

\end{center}

\end{table*}

\begin{table*}[tt]
\caption{ Performance on the state-of-the-art image transformation defense methods }
\label{table:cvprtable04}
\vspace{-.4cm}
\begin{center}
\setlength{\tabcolsep}{1mm}{
\begin{tabular}{|c|c|c|c|c|l|l|l|c|l|l|l|}
\hline
                 Network           &             \diagbox{defender}{attacker}                    & Single-pixel Attack & Boundary Attack & \multicolumn{4}{c|}{CMU(1.5/2/3/4)}       & \multicolumn{4}{c|}{ACMMM2020(1.5/2/3/4)} \\ \hline
\multirow{3}{*}{Resnet101}   & Jpeg defend & 24\%                & 13\%            & \multicolumn{4}{c|}{100\%/98\%/94\%/92\%} & \multicolumn{4}{c|}{97\%/95\%/88\%/83\%} \\ \cline{2-12}
                             & Comdefend   & 17\%                & 13\%            & \multicolumn{4}{c|}{99\%/94\%/88\%/82\%}  & \multicolumn{4}{c|}{97\%/94\%/89\%/82\%} \\ \cline{2-12}
                             & HGD         & 42\%                & 34\%            & \multicolumn{4}{c|}{98\%/95\%/95\%/94\%}  & \multicolumn{4}{c|}{97\%/95\%/92\%/90\%} \\ \hline
\multirow{3}{*}{InceptionV3} & Jpeg defend & 42\%                & 8\%             & \multicolumn{4}{c|}{100\%/97\%/94\%/91\%} & \multicolumn{4}{c|}{99\%/95\%/90\%/87\%} \\ \cline{2-12}
                             & Comdefend   & 34\%                & 12\%            & \multicolumn{4}{c|}{99\%/95\%/91\%/86\%}  & \multicolumn{4}{c|}{98\%/94\%/90\%/86\%} \\ \cline{2-12}
                             & HGD         & 32\%                & 36\%            & \multicolumn{4}{c|}{98\%/95\%/89\%/88\%}  & \multicolumn{4}{c|}{95\%/90\%/86\%/85\%} \\ \hline
\end{tabular}
}

\end{center}

\end{table*}

\subsection{Optimization method implementation}
The initial value of the step size $r$ is set as 0.5. And the initial
$p, q$ and $\alpha$ are set as 0, 0 and 100. The range of the $p$ is
$[0,W_{h}-W_{sw}]$. The range of the $q$ is $[0,H_{h}-H_{sw}]$.
And the range of the $\alpha$ is $[100,200]$.

\subsection{Selection of hyper-parameters}
We conduct a large number of experiments to determine two hyper parameters
in BHE. One is the number of basin hopping iterations $I$, the
other one is crossover probability $\text{CR}$. We adopt BHE to
attack DNN models using ACMMM 2020 logo with scale=1/4. In detail, we compute the
attack success rates of the Resnet101 on 1000 random image of
the ImageNet dataset. The result is shown in Table \ref{table:cvprtable00}.
From Table \ref{table:cvprtable00}, it is clear that the attack success
rate increases when $I$ increases. That is, as the number of Basin Hopping iterations
increases, the solution generated by BH will be better, resulting in achieving
a higher attack success rate. But more iterations mean more time spent.
Considering time complexity, we set $\text{CR}$ to 0.9 and $I$ to 3. In this way, Adv-watermark
can achieve the highest attack success rate($60\%$).
And in the original BH algorithm, the iteration $I$ is set to 450.

\subsection{Attack performance}
\begin{figure}[tt]
\begin{center}
   \includegraphics[width=1.0\linewidth]{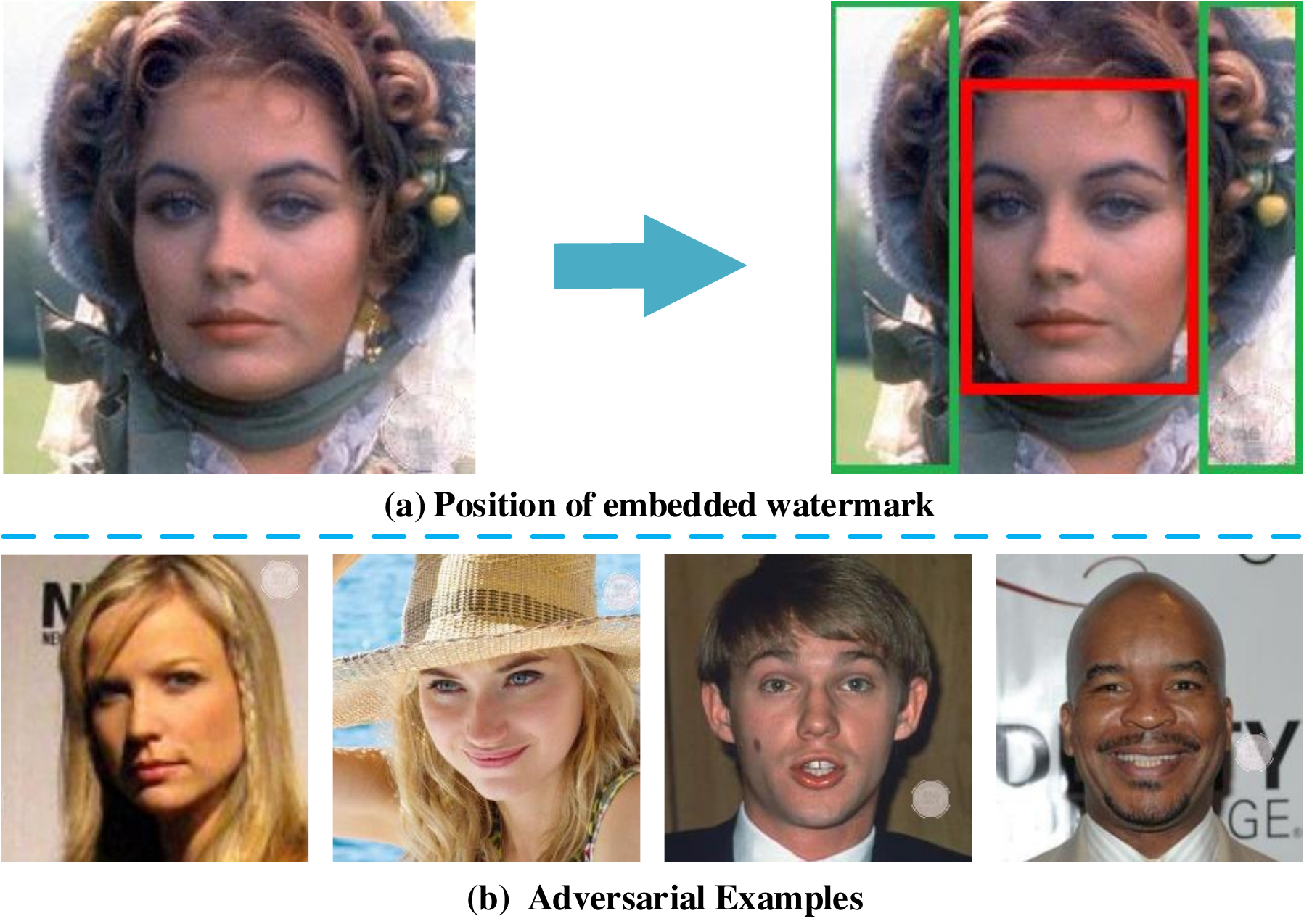}
\end{center}
\caption{ (a) Limit of embedded watermark position. The face is in the red rectangle and the
embedded watermark is restricted to the green rectangles. (b) Adversarial examples on the CASIA-WebFace dataset.}
\label{fig:cvpr6}
\vspace{-.5cm}
\end{figure}

In order to verify the proposed method comprehensively, we
choose five university logo watermarks and five official ACMMM watermarks
as the image watermarks to generate corresponding adversarial examples. And
we also choose five different color fonts as the text watermarks
to generate corresponding adversarial examples. The average attack success rates of individual logos or text watermarks are reported in Table \ref{table:cvprtable01}.
The first column of each row shows the results of BH and the second column of each row shows the results of BHE. It is clear that
the proposed BHE can achieve a high attack success rate.
As for the university logo watermarks, when the watermark size is set as $4/9$ of the host image size,
the attack success rate can achieve about 97\%.
And when the watermark size is set as $1/16$ of the host image size,
the attack rate also can achieve 69\%.
As for the ACMMM logo watermarks, the average attack success rates of them drop a little.
That is because that the height-width ratio
of the ACMMM watermark is not 1:1(the height-width ratios of the
ACMMM logo watermarks(2016-2020) are $1:2.6$, $1:2.5$, $1:3$, $1:2.1$ and $1:2.6$),
and the size of the ACMMM logo watermark is smaller than the
university logo watermark when
the scale is the same. In detail, when scale=$1/4$, the size of ACMMM2018
logo watermark is about $1/48$ of the host watermark size. Even though
the performance of the adversarial ACMMM logo watermarks declines a little,
they also achieve a high attack success rate. Simultaneously, we
use the text watermark to attack the well-trained classification models.
As shown in Table \ref{table:cvprtable01},
the proposed method can achieve about 86\%, 76\%, 81\%, 71\% and 66\% average attack success rates on
Alexnet, VGG19, SqueezeNet, Resnet101 and InceptionV3 with different font sizes.
Compared with BH, the proposed BHE can achieve a higher attack success rate.
Moreover, we conduct a series of experiments on the CASIA-WebFace dataset within
restricting adversarial watermark position.
Specifically, as shown in Figure \ref{fig:cvpr6} (a),
we use MTCNN \cite{DBLP:journals/spl/ZhangZLQ16} to
find face area which is marked as the red rectangle.
We restrict the embedded watermark to the area on
both sides of the rectangle which is marked as the green
rectangle. And then we use Adv-watermark
to attack InceptionV1
which is trained on CASIA-WebFace dataset.
Generated adversarial examples are shown in Figure \ref{fig:cvpr6} (b).
The attack result is shown in Table \ref{table:cvprtable02}. Note that
since we limit the embedding area of the watermark, we should
adopt a smaller scale and font size.
\begin{table}[tt]
\caption{ Performance on the adversarial training }
\label{table:cvprtable05}

\begin{center}
\setlength{\tabcolsep}{1mm}{
\begin{tabular}{|c|c|c|c|c|c|c|}
\hline
\multirow{2}{*}{\begin{tabular}[c]{@{}c@{}}Adversarial \\ Training\end{tabular}} & \multicolumn{2}{c|}{MIT}      & \multicolumn{2}{c|}{ACMMM20}  & \multicolumn{2}{c|}{Red Text} \\ \cline{2-7}
                                                                                 & 1/4           & 1/3           & 1/4           & 1/3           & 28            & 32            \\ \hline
MIT(1/4)                                                                         & \textbf{50\%} & \textbf{55\%} & 74\%          & 80\%          & 91\%          & 92\%          \\ \hline
ACMMM20(1/4)                                                                     & 78\%          & 83\%          & \textbf{43\%} & \textbf{48\%} & 85\%          & 86\%          \\ \hline
Red Text(28)                                                                     & 71\%          & 74\%          & 72\%          & 86\%          & \textbf{44\%} & \textbf{47\%} \\ \hline
\end{tabular}
}
\end{center}
\vspace{-.5cm}
\end{table}

\subsection{Comparisons with other attack methods}
To quantitatively evaluate the proposed method performance,
we compare the proposed method with other black-box attack methods:
spatial attack \cite{engstrom2019exploring}, boundary attack \cite{wiel2017decisionbased},
single-pixel attack \cite{su2019one} and pointwise attack\cite{schott2018adversarially}.
In detail, we choose the SU and ACMMM2017 image watermarks with different scales and
blue font text watermark with the different font sizes to complete the contrast experiments.
Their average attack success rates are shown in Table \ref{table:cvprtable03}.
As shown in Table \ref{table:cvprtable03}, it is clear that
compared with other black-box attack methods,
our attack method can achieve a higher attack success rate.
In particular, the average attack success rate of SU reaches up to
88\%.
\par In order to evaluate the robustness of the proposed method,
we compare the Adv-watermark with other black-box attack methods:
single-pixel attack and boundary attack, and choose three image transformation defense
methods: Jpeg defend \cite{das2017keeping}, Comdefend \cite{jia2019comdefend} and
HGD \cite{liao2018defense}.
From Table \ref{table:cvprtable04}, it is clear that
the existing image transformation defense methods are useful for single-pixel attack and boundary attack,
but not useful for our proposed method. Compared with
other attack methods, the proposed method is more robust.
We also conduct adversarial training \cite{DBLP:conf/iclr/MadryMSTV18} to defend
the proposed attack method. In detail, we inject adversarial examples
generated by MIT, ACMMM2020 image watermark with scale $=1/4$ and red text watermark
with font $=28$ into the original image dataset and retrain three Resnet101
on them respectively. And then we use these watermarks with different sizes
to attack these models. The result is shown in Table \ref{table:cvprtable05}.
It is clear that the adversarial training cannot effectively defend
Adv-watermark. Moreover, using another watermark to attack the
adversarial training model can achieve a higher attack
success rate. In other words, even though adversarial training increases the robustness to one
watermark perturbation, it increases the vulnerability to another
watermark perturbation.

\begin{figure}[tt]
\begin{center}
   \includegraphics[width=1\linewidth]{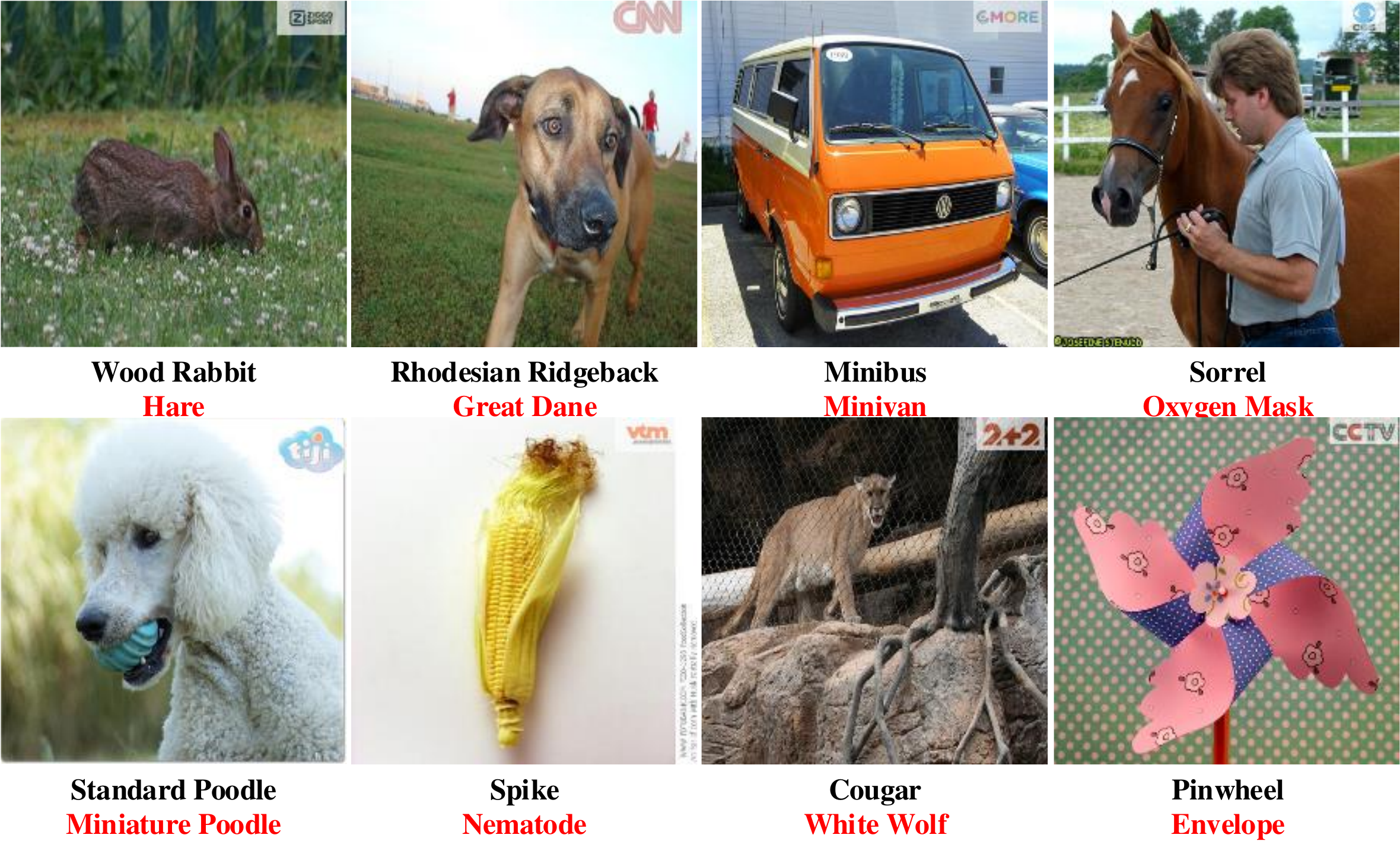}
\end{center}
   \caption{ The adversarial examples with a variety of TV station logos.
   The original class labels are in black color
and the class labels of the adversarial examples are in red color.}
\label{fig:cvpr7}
\vspace{-.3cm}
\end{figure}

\begin{figure}[tt]
\begin{center}
   \includegraphics[width=0.8\linewidth]{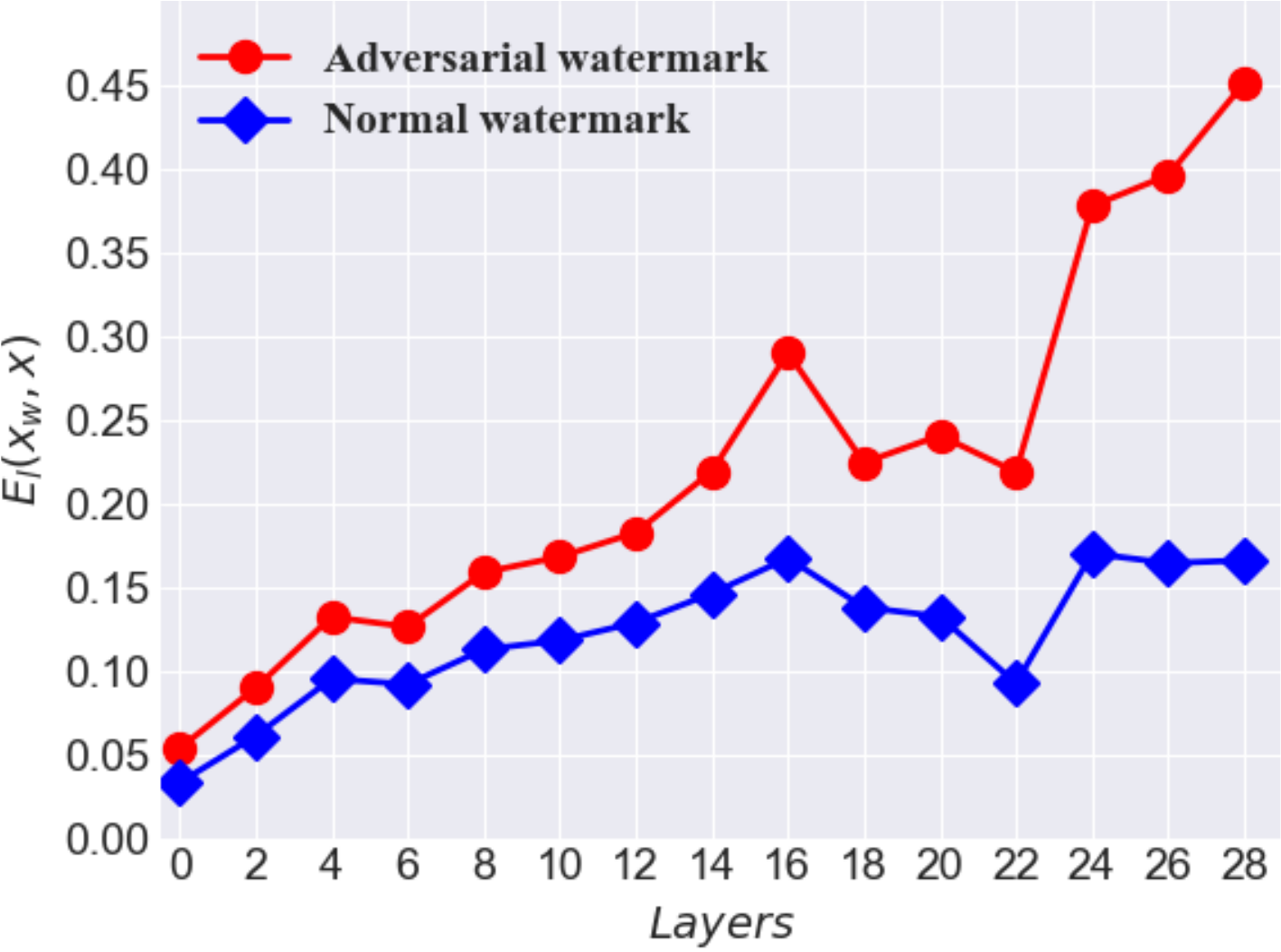}
\end{center}
\caption{ Layer-wise perturbation levels of the VGG16 model.
         Adversarial watermark and normal watermark added to clean images
         correspond to the $E_{l}$, respectively.}
\label{fig:final}
\vspace{-.4cm}
\end{figure}

\subsection{Extension}
The proposed method is not limited to using a watermark to generate
an adversarial example. It can be extended to use the TV station logos
to complete the attack. To make the generated adversarial examples
 more realistic and imperceptible, we also choose more
commonly used TV station logos to complete the attack.
In detail, we select a variety of TV station logos, next
limit the embedded position of the logos to the upper
right corner of the host image and then use the proposed method
to generate the adversarial examples. As shown in Figure \ref{fig:cvpr7}, the generated
adversarial examples are more realistic and common in the physical world.

\subsection{Analysis for Adv-watermark}
Compared with the previous attack methods, Adv-watermark pays more attention to generate realistic adversarial examples. We find DNN models are spatially vulnerable, which adding perturbations at a specific position to clean images can attack them easily. To investigate this characteristic,
we conduct a comparative experiment to evaluate layer-wise perturbations
of the VGG16 model fed adversarial watermark images and normal watermark
images, respectively.  The difference between normal watermarks and adversarial
watermarks is that they are positioned differently on clean images.                                                                                                                                                                                                                                                                                          The perturbation level in layer $l$ can be formulated as:
\begin{equation}
E_{l}\left(x_{w}, x\right)=\left\|f_{l}\left(x_{w}\right)-f_{l}(x)\right\|_{2} /\left\|f_{l}(x)\right\|_{2},
\end{equation}
where $x$ represents a clean image, $x_{w}$ represents the clean image with
adversarial or normal watermark and $f_{l}(\cdot)$ represents the $l$-th
layer of the VGG16 model.
\par The result is shown in Figure \ref{fig:final}. The red curve represents
the $E_{l}$ for adversarial watermark perturbations and the blue curve
represents the $E_{l}$ for normal watermark perturbations.
Specifically, the red curve is the average result on 30 randomly picked images with the adversarial
watermarks and the blue curve is the average result on 30 same images with
the normal watermarks. It is clear that the watermark perturbation is progressively
enlarged with the layer hierarchy. But in the top layer, the adversarial watermark perturbation is
much higher than the normal watermark perturbation. Because the classification result
is dependent on the top-level features, the adversarial watermark perturbation can fool
DNN models but the normal watermark perturbation can not.

\section{Conclusion}
In this paper, we discovered DNN models were spatially vulnerable,
which adding perturbations at a specific position to clean images could attack models easily.
And then we proposed a novel attacking method which used the real watermark
to attack the well-trained classifier. Our adversarial perturbation was meaningful, which was different from the traditional ones. We formulated the watermark attack problem as a global optimization problem, and
proposed a novel optimization
algorithm(BHE) to generate adversarial examples. Compared with
the previous BH, BHE achieved a higher attack success rate. Moreover, the Adv-watermark was more robust,
because the image transformation defense methods could not defend the proposed attack
method. And the proposed method
could be more commonly used in the real world.
\section*{Acknowledgement}
Supported by the National Key R\&D Program of China under Grant 2018AAA0102503, National Natural Science Foundation of China (No. U1936210, U1736219, 61971016, 61806109), The Open Research Fund from Shenzhen Research Institute of Big Data, under Grant No. 2019ORF01010, Beijing Natural Science Foundation (No.L182057), Peng Cheng Laboratory Project of Guangdong Province PCL2018KP004.

\par\vfill\par

\clearpage

\bibliographystyle{ACM-Reference-Format}
\bibliography{sample-base}


\begin{thebibliography}{38}


\ifx \showCODEN    \undefined \def \showCODEN     #1{\unskip}     \fi
\ifx \showDOI      \undefined \def \showDOI       #1{#1}\fi
\ifx \showISBNx    \undefined \def \showISBNx     #1{\unskip}     \fi
\ifx \showISBNxiii \undefined \def \showISBNxiii  #1{\unskip}     \fi
\ifx \showISSN     \undefined \def \showISSN      #1{\unskip}     \fi
\ifx \showLCCN     \undefined \def \showLCCN      #1{\unskip}     \fi
\ifx \shownote     \undefined \def \shownote      #1{#1}          \fi
\ifx \showarticletitle \undefined \def \showarticletitle #1{#1}   \fi
\ifx \showURL      \undefined \def \showURL       {\relax}        \fi
\providecommand\bibfield[2]{#2}
\providecommand\bibinfo[2]{#2}
\providecommand\natexlab[1]{#1}
\providecommand\showeprint[2][]{arXiv:#2}

\bibitem[\protect\citeauthoryear{Brendel, Rauber, and Bethge}{Brendel
  et~al\mbox{.}}{2017}]%
        {wiel2017decisionbased}
\bibfield{author}{\bibinfo{person}{Wieland Brendel}, \bibinfo{person}{Jonas
  Rauber}, {and} \bibinfo{person}{Matthias Bethge}.}
  \bibinfo{year}{2017}\natexlab{}.
\newblock \bibinfo{title}{Decision-Based Adversarial Attacks: Reliable Attacks
  Against Black-Box Machine Learning Models}.
\newblock
\newblock
\showeprint[arxiv]{1712.04248}~[stat.ML]


\bibitem[\protect\citeauthoryear{Carlini and Wagner}{Carlini and
  Wagner}{2017}]%
        {carlini2017towards}
\bibfield{author}{\bibinfo{person}{Nicholas Carlini} {and}
  \bibinfo{person}{David Wagner}.} \bibinfo{year}{2017}\natexlab{}.
\newblock \showarticletitle{Towards evaluating the robustness of neural
  networks}. In \bibinfo{booktitle}{\emph{2017 IEEE Symposium on Security and
  Privacy (SP)}}. IEEE, \bibinfo{pages}{39--57}.
\newblock


\bibitem[\protect\citeauthoryear{Das, Shanbhogue, Chen, Hohman, Chen, Kounavis,
  and Chau}{Das et~al\mbox{.}}{2017}]%
        {das2017keeping}
\bibfield{author}{\bibinfo{person}{Nilaksh Das}, \bibinfo{person}{Madhuri
  Shanbhogue}, \bibinfo{person}{Shang-Tse Chen}, \bibinfo{person}{Fred Hohman},
  \bibinfo{person}{Li Chen}, \bibinfo{person}{Michael~E Kounavis}, {and}
  \bibinfo{person}{Duen~Horng Chau}.} \bibinfo{year}{2017}\natexlab{}.
\newblock \showarticletitle{Keeping the bad guys out: Protecting and
  vaccinating deep learning with jpeg compression}.
\newblock \bibinfo{journal}{\emph{arXiv preprint arXiv:1705.02900}}
  (\bibinfo{year}{2017}).
\newblock


\bibitem[\protect\citeauthoryear{Engstrom, Tran, Tsipras, Schmidt, and
  Madry}{Engstrom et~al\mbox{.}}{2019}]%
        {engstrom2019exploring}
\bibfield{author}{\bibinfo{person}{Logan Engstrom}, \bibinfo{person}{Brandon
  Tran}, \bibinfo{person}{Dimitris Tsipras}, \bibinfo{person}{Ludwig Schmidt},
  {and} \bibinfo{person}{Aleksander Madry}.} \bibinfo{year}{2019}\natexlab{}.
\newblock \showarticletitle{Exploring the Landscape of Spatial Robustness}. In
  \bibinfo{booktitle}{\emph{International Conference on Machine Learning}}.
  \bibinfo{pages}{1802--1811}.
\newblock


\bibitem[\protect\citeauthoryear{Goodfellow, Pouget-Abadie, Mirza, Xu,
  Warde-Farley, Ozair, Courville, and Bengio}{Goodfellow
  et~al\mbox{.}}{2014a}]%
        {goodfellow2014generative}
\bibfield{author}{\bibinfo{person}{Ian Goodfellow}, \bibinfo{person}{Jean
  Pouget-Abadie}, \bibinfo{person}{Mehdi Mirza}, \bibinfo{person}{Bing Xu},
  \bibinfo{person}{David Warde-Farley}, \bibinfo{person}{Sherjil Ozair},
  \bibinfo{person}{Aaron Courville}, {and} \bibinfo{person}{Yoshua Bengio}.}
  \bibinfo{year}{2014}\natexlab{a}.
\newblock \showarticletitle{Generative adversarial nets}. In
  \bibinfo{booktitle}{\emph{Advances in neural information processing
  systems}}. \bibinfo{pages}{2672--2680}.
\newblock


\bibitem[\protect\citeauthoryear{Goodfellow, Shlens, and Szegedy}{Goodfellow
  et~al\mbox{.}}{2014b}]%
        {goodfellow2014explaining}
\bibfield{author}{\bibinfo{person}{Ian~J Goodfellow}, \bibinfo{person}{Jonathon
  Shlens}, {and} \bibinfo{person}{Christian Szegedy}.}
  \bibinfo{year}{2014}\natexlab{b}.
\newblock \showarticletitle{Explaining and harnessing adversarial examples}.
\newblock \bibinfo{journal}{\emph{arXiv preprint arXiv:1412.6572}}
  (\bibinfo{year}{2014}).
\newblock


\bibitem[\protect\citeauthoryear{Hartung and Kutter}{Hartung and
  Kutter}{1999}]%
        {hartung1999multimedia}
\bibfield{author}{\bibinfo{person}{Frank Hartung} {and} \bibinfo{person}{Martin
  Kutter}.} \bibinfo{year}{1999}\natexlab{}.
\newblock \showarticletitle{Multimedia watermarking techniques}.
\newblock \bibinfo{journal}{\emph{Proc. IEEE}} \bibinfo{volume}{87},
  \bibinfo{number}{7} (\bibinfo{year}{1999}), \bibinfo{pages}{1079--1107}.
\newblock


\bibitem[\protect\citeauthoryear{He, Zhang, Ren, and Sun}{He
  et~al\mbox{.}}{2016}]%
        {he2016deep}
\bibfield{author}{\bibinfo{person}{Kaiming He}, \bibinfo{person}{Xiangyu
  Zhang}, \bibinfo{person}{Shaoqing Ren}, {and} \bibinfo{person}{Jian Sun}.}
  \bibinfo{year}{2016}\natexlab{}.
\newblock \showarticletitle{Deep residual learning for image recognition}. In
  \bibinfo{booktitle}{\emph{Proceedings of the IEEE conference on computer
  vision and pattern recognition}}. \bibinfo{pages}{770--778}.
\newblock


\bibitem[\protect\citeauthoryear{Hu, Kwong, and Huang}{Hu
  et~al\mbox{.}}{2005}]%
        {hu2005algorithm}
\bibfield{author}{\bibinfo{person}{Yongjian Hu}, \bibinfo{person}{Sam Kwong},
  {and} \bibinfo{person}{Jiwu Huang}.} \bibinfo{year}{2005}\natexlab{}.
\newblock \showarticletitle{An algorithm for removable visible watermarking}.
\newblock \bibinfo{journal}{\emph{IEEE Transactions on Circuits and Systems for
  Video Technology}} \bibinfo{volume}{16}, \bibinfo{number}{1}
  (\bibinfo{year}{2005}), \bibinfo{pages}{129--133}.
\newblock


\bibitem[\protect\citeauthoryear{Huang and Tang}{Huang and Tang}{2006}]%
        {huang2006contrast}
\bibfield{author}{\bibinfo{person}{Biao-Bing Huang} {and}
  \bibinfo{person}{Shao-Xian Tang}.} \bibinfo{year}{2006}\natexlab{}.
\newblock \showarticletitle{A contrast-sensitive visible watermarking scheme}.
\newblock \bibinfo{journal}{\emph{IEEE MultiMedia}} \bibinfo{volume}{13},
  \bibinfo{number}{2} (\bibinfo{year}{2006}), \bibinfo{pages}{60--66}.
\newblock


\bibitem[\protect\citeauthoryear{Iandola, Han, Moskewicz, Ashraf, Dally, and
  Keutzer}{Iandola et~al\mbox{.}}{2016}]%
        {iandola2016squeezenet}
\bibfield{author}{\bibinfo{person}{Forrest~N Iandola}, \bibinfo{person}{Song
  Han}, \bibinfo{person}{Matthew~W Moskewicz}, \bibinfo{person}{Khalid Ashraf},
  \bibinfo{person}{William~J Dally}, {and} \bibinfo{person}{Kurt Keutzer}.}
  \bibinfo{year}{2016}\natexlab{}.
\newblock \showarticletitle{SqueezeNet: AlexNet-level accuracy with 50x fewer
  parameters and< 0.5 MB model size}.
\newblock \bibinfo{journal}{\emph{arXiv preprint arXiv:1602.07360}}
  (\bibinfo{year}{2016}).
\newblock


\bibitem[\protect\citeauthoryear{Jia, Wei, Cao, and Foroosh}{Jia
  et~al\mbox{.}}{2019}]%
        {jia2019comdefend}
\bibfield{author}{\bibinfo{person}{Xiaojun Jia}, \bibinfo{person}{Xingxing
  Wei}, \bibinfo{person}{Xiaochun Cao}, {and} \bibinfo{person}{Hassan
  Foroosh}.} \bibinfo{year}{2019}\natexlab{}.
\newblock \showarticletitle{ComDefend: An Efficient Image Compression Model to
  Defend Adversarial Examples}. In \bibinfo{booktitle}{\emph{Proceedings of the
  IEEE Conference on Computer Vision and Pattern Recognition}}.
  \bibinfo{pages}{6084--6092}.
\newblock


\bibitem[\protect\citeauthoryear{Kankanhalli, Ramakrishnan,
  et~al\mbox{.}}{Kankanhalli et~al\mbox{.}}{1999}]%
        {kankanhalli1999adaptive}
\bibfield{author}{\bibinfo{person}{Mohan~S Kankanhalli}, \bibinfo{person}{KR
  Ramakrishnan}, {et~al\mbox{.}}} \bibinfo{year}{1999}\natexlab{}.
\newblock \showarticletitle{Adaptive visible watermarking of images}. In
  \bibinfo{booktitle}{\emph{Proceedings IEEE International Conference on
  Multimedia Computing and Systems}}, Vol.~\bibinfo{volume}{1}. IEEE,
  \bibinfo{pages}{568--573}.
\newblock


\bibitem[\protect\citeauthoryear{Krizhevsky, Sutskever, and Hinton}{Krizhevsky
  et~al\mbox{.}}{2012}]%
        {krizhevsky2012imagenet}
\bibfield{author}{\bibinfo{person}{Alex Krizhevsky}, \bibinfo{person}{Ilya
  Sutskever}, {and} \bibinfo{person}{Geoffrey~E Hinton}.}
  \bibinfo{year}{2012}\natexlab{}.
\newblock \showarticletitle{Imagenet classification with deep convolutional
  neural networks}. In \bibinfo{booktitle}{\emph{Advances in neural information
  processing systems}}. \bibinfo{pages}{1097--1105}.
\newblock


\bibitem[\protect\citeauthoryear{Kurakin, Goodfellow, and Bengio}{Kurakin
  et~al\mbox{.}}{2016}]%
        {kurakin2016adversarial}
\bibfield{author}{\bibinfo{person}{Alexey Kurakin}, \bibinfo{person}{Ian
  Goodfellow}, {and} \bibinfo{person}{Samy Bengio}.}
  \bibinfo{year}{2016}\natexlab{}.
\newblock \showarticletitle{Adversarial examples in the physical world}.
\newblock \bibinfo{journal}{\emph{arXiv preprint arXiv:1607.02533}}
  (\bibinfo{year}{2016}).
\newblock


\bibitem[\protect\citeauthoryear{Lagae, Lefebvre, Cook, DeRose, Drettakis,
  Ebert, Lewis, Perlin, and Zwicker}{Lagae et~al\mbox{.}}{2010}]%
        {lagae2010survey}
\bibfield{author}{\bibinfo{person}{Ares Lagae}, \bibinfo{person}{Sylvain
  Lefebvre}, \bibinfo{person}{Rob Cook}, \bibinfo{person}{Tony DeRose},
  \bibinfo{person}{George Drettakis}, \bibinfo{person}{David~S Ebert},
  \bibinfo{person}{John~P Lewis}, \bibinfo{person}{Ken Perlin}, {and}
  \bibinfo{person}{Matthias Zwicker}.} \bibinfo{year}{2010}\natexlab{}.
\newblock \showarticletitle{A survey of procedural noise functions}. In
  \bibinfo{booktitle}{\emph{Computer Graphics Forum}},
  Vol.~\bibinfo{volume}{29}. Wiley Online Library, \bibinfo{pages}{2579--2600}.
\newblock


\bibitem[\protect\citeauthoryear{Liang, Wei, Yao, and Cao}{Liang
  et~al\mbox{.}}{2020}]%
        {liang2020efficient}
\bibfield{author}{\bibinfo{person}{Siyuan Liang}, \bibinfo{person}{Xingxing
  Wei}, \bibinfo{person}{Siyuan Yao}, {and} \bibinfo{person}{Xiaochun Cao}.}
  \bibinfo{year}{2020}\natexlab{}.
\newblock \showarticletitle{Efficient Adversarial Attacks for Visual Object
  Tracking}.
\newblock \bibinfo{journal}{\emph{arXiv preprint arXiv:2008.00217}}
  (\bibinfo{year}{2020}).
\newblock


\bibitem[\protect\citeauthoryear{Liao, Liang, Dong, Pang, Hu, and Zhu}{Liao
  et~al\mbox{.}}{2018}]%
        {liao2018defense}
\bibfield{author}{\bibinfo{person}{Fangzhou Liao}, \bibinfo{person}{Ming
  Liang}, \bibinfo{person}{Yinpeng Dong}, \bibinfo{person}{Tianyu Pang},
  \bibinfo{person}{Xiaolin Hu}, {and} \bibinfo{person}{Jun Zhu}.}
  \bibinfo{year}{2018}\natexlab{}.
\newblock \showarticletitle{Defense against adversarial attacks using
  high-level representation guided denoiser}. In
  \bibinfo{booktitle}{\emph{Proceedings of the IEEE Conference on Computer
  Vision and Pattern Recognition}}. \bibinfo{pages}{1778--1787}.
\newblock


\bibitem[\protect\citeauthoryear{Liu and Tsai}{Liu and Tsai}{2010}]%
        {liu2010generic}
\bibfield{author}{\bibinfo{person}{Tsung-Yuan Liu} {and}
  \bibinfo{person}{Wen-Hsiang Tsai}.} \bibinfo{year}{2010}\natexlab{}.
\newblock \showarticletitle{Generic lossless visible watermarking—a new
  approach}.
\newblock \bibinfo{journal}{\emph{IEEE transactions on image processing}}
  \bibinfo{volume}{19}, \bibinfo{number}{5} (\bibinfo{year}{2010}),
  \bibinfo{pages}{1224--1235}.
\newblock


\bibitem[\protect\citeauthoryear{Madry, Makelov, Schmidt, Tsipras, and
  Vladu}{Madry et~al\mbox{.}}{2018}]%
        {DBLP:conf/iclr/MadryMSTV18}
\bibfield{author}{\bibinfo{person}{Aleksander Madry},
  \bibinfo{person}{Aleksandar Makelov}, \bibinfo{person}{Ludwig Schmidt},
  \bibinfo{person}{Dimitris Tsipras}, {and} \bibinfo{person}{Adrian Vladu}.}
  \bibinfo{year}{2018}\natexlab{}.
\newblock \showarticletitle{Towards Deep Learning Models Resistant to
  Adversarial Attacks}. In \bibinfo{booktitle}{\emph{{ICLR} (Poster)}}.
  \bibinfo{publisher}{OpenReview.net}.
\newblock


\bibitem[\protect\citeauthoryear{Mintzer, Braudaway, and Yeung}{Mintzer
  et~al\mbox{.}}{1997}]%
        {mintzer1997effective}
\bibfield{author}{\bibinfo{person}{Fred Mintzer}, \bibinfo{person}{Gordon~W
  Braudaway}, {and} \bibinfo{person}{Minerva~M Yeung}.}
  \bibinfo{year}{1997}\natexlab{}.
\newblock \showarticletitle{Effective and ineffective digital watermarks}. In
  \bibinfo{booktitle}{\emph{Proceedings of International Conference on Image
  Processing}}, Vol.~\bibinfo{volume}{3}. IEEE, \bibinfo{pages}{9--12}.
\newblock


\bibitem[\protect\citeauthoryear{Moosavi-Dezfooli, Fawzi, Fawzi, and
  Frossard}{Moosavi-Dezfooli et~al\mbox{.}}{2017}]%
        {moosavi2017universal}
\bibfield{author}{\bibinfo{person}{Seyed-Mohsen Moosavi-Dezfooli},
  \bibinfo{person}{Alhussein Fawzi}, \bibinfo{person}{Omar Fawzi}, {and}
  \bibinfo{person}{Pascal Frossard}.} \bibinfo{year}{2017}\natexlab{}.
\newblock \showarticletitle{Universal adversarial perturbations}. In
  \bibinfo{booktitle}{\emph{Proceedings of the IEEE conference on computer
  vision and pattern recognition}}. \bibinfo{pages}{1765--1773}.
\newblock


\bibitem[\protect\citeauthoryear{Moosavi-Dezfooli, Fawzi, and
  Frossard}{Moosavi-Dezfooli et~al\mbox{.}}{2016}]%
        {moosavi2016deepfool}
\bibfield{author}{\bibinfo{person}{Seyed-Mohsen Moosavi-Dezfooli},
  \bibinfo{person}{Alhussein Fawzi}, {and} \bibinfo{person}{Pascal Frossard}.}
  \bibinfo{year}{2016}\natexlab{}.
\newblock \showarticletitle{Deepfool: a simple and accurate method to fool deep
  neural networks}. In \bibinfo{booktitle}{\emph{Proceedings of the IEEE
  conference on computer vision and pattern recognition}}.
  \bibinfo{pages}{2574--2582}.
\newblock


\bibitem[\protect\citeauthoryear{Papernot, McDaniel, Jha, Fredrikson, Celik,
  and Swami}{Papernot et~al\mbox{.}}{2016a}]%
        {papernot2016limitations}
\bibfield{author}{\bibinfo{person}{Nicolas Papernot}, \bibinfo{person}{Patrick
  McDaniel}, \bibinfo{person}{Somesh Jha}, \bibinfo{person}{Matt Fredrikson},
  \bibinfo{person}{Z~Berkay Celik}, {and} \bibinfo{person}{Ananthram Swami}.}
  \bibinfo{year}{2016}\natexlab{a}.
\newblock \showarticletitle{The limitations of deep learning in adversarial
  settings}. In \bibinfo{booktitle}{\emph{2016 IEEE European Symposium on
  Security and Privacy (EuroS\&P)}}. IEEE, \bibinfo{pages}{372--387}.
\newblock


\bibitem[\protect\citeauthoryear{Papernot, McDaniel, Wu, Jha, and
  Swami}{Papernot et~al\mbox{.}}{2016b}]%
        {papernot2016distillation}
\bibfield{author}{\bibinfo{person}{Nicolas Papernot}, \bibinfo{person}{Patrick
  McDaniel}, \bibinfo{person}{Xi Wu}, \bibinfo{person}{Somesh Jha}, {and}
  \bibinfo{person}{Ananthram Swami}.} \bibinfo{year}{2016}\natexlab{b}.
\newblock \showarticletitle{Distillation as a defense to adversarial
  perturbations against deep neural networks}. In
  \bibinfo{booktitle}{\emph{2016 IEEE Symposium on Security and Privacy (SP)}}.
  IEEE, \bibinfo{pages}{582--597}.
\newblock


\bibitem[\protect\citeauthoryear{Qiu, Xiao, Yang, Yan, Lee, and Li}{Qiu
  et~al\mbox{.}}{2019}]%
        {qiu2019semanticadv}
\bibfield{author}{\bibinfo{person}{Haonan Qiu}, \bibinfo{person}{Chaowei Xiao},
  \bibinfo{person}{Lei Yang}, \bibinfo{person}{Xinchen Yan},
  \bibinfo{person}{Honglak Lee}, {and} \bibinfo{person}{Bo Li}.}
  \bibinfo{year}{2019}\natexlab{}.
\newblock \showarticletitle{SemanticAdv: Generating Adversarial Examples via
  Attribute-conditional Image Editing}.
\newblock \bibinfo{journal}{\emph{arXiv preprint arXiv:1906.07927}}
  (\bibinfo{year}{2019}).
\newblock


\bibitem[\protect\citeauthoryear{Rauber, Brendel, and Bethge}{Rauber
  et~al\mbox{.}}{2017}]%
        {rauber2017foolbox}
\bibfield{author}{\bibinfo{person}{Jonas Rauber}, \bibinfo{person}{Wieland
  Brendel}, {and} \bibinfo{person}{Matthias Bethge}.}
  \bibinfo{year}{2017}\natexlab{}.
\newblock \showarticletitle{Foolbox: A Python toolbox to benchmark the
  robustness of machine learning models}. In \bibinfo{booktitle}{\emph{Reliable
  Machine Learning in the Wild Workshop, 34th International Conference on
  Machine Learning}}.
\newblock
\urldef\tempurl%
\url{http://arxiv.org/abs/1707.04131}
\showURL{%
\tempurl}


\bibitem[\protect\citeauthoryear{Russakovsky, Deng, Su, Krause, Satheesh, Ma,
  Huang, Karpathy, Khosla, Bernstein, et~al\mbox{.}}{Russakovsky
  et~al\mbox{.}}{2015}]%
        {russakovsky2015imagenet}
\bibfield{author}{\bibinfo{person}{Olga Russakovsky}, \bibinfo{person}{Jia
  Deng}, \bibinfo{person}{Hao Su}, \bibinfo{person}{Jonathan Krause},
  \bibinfo{person}{Sanjeev Satheesh}, \bibinfo{person}{Sean Ma},
  \bibinfo{person}{Zhiheng Huang}, \bibinfo{person}{Andrej Karpathy},
  \bibinfo{person}{Aditya Khosla}, \bibinfo{person}{Michael Bernstein},
  {et~al\mbox{.}}} \bibinfo{year}{2015}\natexlab{}.
\newblock \showarticletitle{Imagenet large scale visual recognition challenge}.
\newblock \bibinfo{journal}{\emph{International journal of computer vision}}
  \bibinfo{volume}{115}, \bibinfo{number}{3} (\bibinfo{year}{2015}),
  \bibinfo{pages}{211--252}.
\newblock


\bibitem[\protect\citeauthoryear{Schott, Rauber, Bethge, and Brendel}{Schott
  et~al\mbox{.}}{2018}]%
        {schott2018adversarially}
\bibfield{author}{\bibinfo{person}{Lukas Schott}, \bibinfo{person}{Jonas
  Rauber}, \bibinfo{person}{Matthias Bethge}, {and} \bibinfo{person}{Wieland
  Brendel}.} \bibinfo{year}{2018}\natexlab{}.
\newblock \bibinfo{title}{Towards the first adversarially robust neural network
  model on MNIST}.
\newblock
\newblock
\showeprint[arxiv]{1805.09190}~[cs.CV]


\bibitem[\protect\citeauthoryear{Selvaraju, Cogswell, Das, Vedantam, Parikh,
  and Batra}{Selvaraju et~al\mbox{.}}{2017}]%
        {Selvaraju_2017_ICCV}
\bibfield{author}{\bibinfo{person}{Ramprasaath~R. Selvaraju},
  \bibinfo{person}{Michael Cogswell}, \bibinfo{person}{Abhishek Das},
  \bibinfo{person}{Ramakrishna Vedantam}, \bibinfo{person}{Devi Parikh}, {and}
  \bibinfo{person}{Dhruv Batra}.} \bibinfo{year}{2017}\natexlab{}.
\newblock \showarticletitle{Grad-CAM: Visual Explanations From Deep Networks
  via Gradient-Based Localization}. In \bibinfo{booktitle}{\emph{The IEEE
  International Conference on Computer Vision (ICCV)}}.
\newblock


\bibitem[\protect\citeauthoryear{Shen, Sethi, and Bhaskaran}{Shen
  et~al\mbox{.}}{1998}]%
        {shen1998dct}
\bibfield{author}{\bibinfo{person}{Bo Shen}, \bibinfo{person}{Ishwar~K Sethi},
  {and} \bibinfo{person}{Vasudev Bhaskaran}.} \bibinfo{year}{1998}\natexlab{}.
\newblock \showarticletitle{DCT domain alpha blending}. In
  \bibinfo{booktitle}{\emph{Proceedings 1998 International Conference on Image
  Processing. ICIP98 (Cat. No. 98CB36269)}}, Vol.~\bibinfo{volume}{1}. IEEE,
  \bibinfo{pages}{857--861}.
\newblock


\bibitem[\protect\citeauthoryear{Simonyan and Zisserman}{Simonyan and
  Zisserman}{2014}]%
        {simonyan2014very}
\bibfield{author}{\bibinfo{person}{Karen Simonyan} {and}
  \bibinfo{person}{Andrew Zisserman}.} \bibinfo{year}{2014}\natexlab{}.
\newblock \showarticletitle{Very deep convolutional networks for large-scale
  image recognition}.
\newblock \bibinfo{journal}{\emph{arXiv preprint arXiv:1409.1556}}
  (\bibinfo{year}{2014}).
\newblock


\bibitem[\protect\citeauthoryear{Su, Vargas, and Sakurai}{Su
  et~al\mbox{.}}{2019}]%
        {su2019one}
\bibfield{author}{\bibinfo{person}{Jiawei Su},
  \bibinfo{person}{Danilo~Vasconcellos Vargas}, {and} \bibinfo{person}{Kouichi
  Sakurai}.} \bibinfo{year}{2019}\natexlab{}.
\newblock \showarticletitle{One pixel attack for fooling deep neural networks}.
\newblock \bibinfo{journal}{\emph{IEEE Transactions on Evolutionary
  Computation}} (\bibinfo{year}{2019}).
\newblock


\bibitem[\protect\citeauthoryear{Szegedy, Liu, Jia, Sermanet, Reed, Anguelov,
  Erhan, Vanhoucke, and Rabinovich}{Szegedy et~al\mbox{.}}{2015}]%
        {DBLP:conf/cvpr/SzegedyLJSRAEVR15}
\bibfield{author}{\bibinfo{person}{Christian Szegedy}, \bibinfo{person}{Wei
  Liu}, \bibinfo{person}{Yangqing Jia}, \bibinfo{person}{Pierre Sermanet},
  \bibinfo{person}{Scott~E. Reed}, \bibinfo{person}{Dragomir Anguelov},
  \bibinfo{person}{Dumitru Erhan}, \bibinfo{person}{Vincent Vanhoucke}, {and}
  \bibinfo{person}{Andrew Rabinovich}.} \bibinfo{year}{2015}\natexlab{}.
\newblock \showarticletitle{Going deeper with convolutions}. In
  \bibinfo{booktitle}{\emph{{CVPR}}}. \bibinfo{publisher}{{IEEE} Computer
  Society}, \bibinfo{pages}{1--9}.
\newblock


\bibitem[\protect\citeauthoryear{Szegedy, Vanhoucke, Ioffe, Shlens, and
  Wojna}{Szegedy et~al\mbox{.}}{2016}]%
        {szegedy2016rethinking}
\bibfield{author}{\bibinfo{person}{Christian Szegedy}, \bibinfo{person}{Vincent
  Vanhoucke}, \bibinfo{person}{Sergey Ioffe}, \bibinfo{person}{Jon Shlens},
  {and} \bibinfo{person}{Zbigniew Wojna}.} \bibinfo{year}{2016}\natexlab{}.
\newblock \showarticletitle{Rethinking the inception architecture for computer
  vision}. In \bibinfo{booktitle}{\emph{Proceedings of the IEEE conference on
  computer vision and pattern recognition}}. \bibinfo{pages}{2818--2826}.
\newblock


\bibitem[\protect\citeauthoryear{Wales and Doye}{Wales and Doye}{1997}]%
        {wales1997global}
\bibfield{author}{\bibinfo{person}{David~J Wales} {and}
  \bibinfo{person}{Jonathan P~K Doye}.} \bibinfo{year}{1997}\natexlab{}.
\newblock \showarticletitle{Global Optimization by Basin-Hopping and the Lowest
  Energy Structures of Lennard-Jones Clusters Containing up to 110 Atoms}.
\newblock \bibinfo{journal}{\emph{Journal of Physical Chemistry A}}
  \bibinfo{volume}{101}, \bibinfo{number}{28} (\bibinfo{year}{1997}),
  \bibinfo{pages}{5111--5116}.
\newblock


\bibitem[\protect\citeauthoryear{Yi, Lei, Liao, and Li}{Yi
  et~al\mbox{.}}{2014}]%
        {yi2014learning}
\bibfield{author}{\bibinfo{person}{Dong Yi}, \bibinfo{person}{Zhen Lei},
  \bibinfo{person}{Shengcai Liao}, {and} \bibinfo{person}{Stan~Z Li}.}
  \bibinfo{year}{2014}\natexlab{}.
\newblock \showarticletitle{Learning face representation from scratch}.
\newblock \bibinfo{journal}{\emph{arXiv preprint arXiv:1411.7923}}
  (\bibinfo{year}{2014}).
\newblock


\bibitem[\protect\citeauthoryear{Zhang, Zhang, Li, and Qiao}{Zhang
  et~al\mbox{.}}{2016}]%
        {DBLP:journals/spl/ZhangZLQ16}
\bibfield{author}{\bibinfo{person}{Kaipeng Zhang}, \bibinfo{person}{Zhanpeng
  Zhang}, \bibinfo{person}{Zhifeng Li}, {and} \bibinfo{person}{Yu Qiao}.}
  \bibinfo{year}{2016}\natexlab{}.
\newblock \showarticletitle{Joint Face Detection and Alignment Using Multitask
  Cascaded Convolutional Networks}.
\newblock \bibinfo{journal}{\emph{{IEEE} Signal Process. Lett.}}
  \bibinfo{volume}{23}, \bibinfo{number}{10} (\bibinfo{year}{2016}),
  \bibinfo{pages}{1499--1503}.
\newblock


\end{thebibliography}

\end{document}